\documentclass[12pt]{article}
\usepackage{amsmath, amssymb, amsfonts, amsthm, mathabx, bm, bbm}
\usepackage[dvipsnames]{xcolor}
\usepackage{booktabs}
\allowdisplaybreaks
\usepackage{graphicx,psfrag,epsf}
\usepackage{enumerate}
\usepackage[shortlabels]{enumitem}
\usepackage[round, authoryear]{natbib}
\renewcommand{\citet}{\cite}
\usepackage{url} 
\usepackage{adjustbox}
\usepackage{multicol, multirow}
\usepackage{tabularx}
\newcommand{\bigcell}[2]{\begin{tabular}{@{}#1@{}}#2\end{tabular}}
\newcommand{\blind}{0}

\usepackage{algorithm}
\usepackage{algpseudocode}
\usepackage{lipsum}


\usepackage{xr}
\makeatletter
\newcommand*{\addFileDependency}[1]{
  \typeout{(#1)}
  \@addtofilelist{#1}
  \IfFileExists{#1}{}{\typeout{No file #1.}}
}
\makeatother

\usepackage{tablefootnote}

\usepackage[colorlinks=true, citecolor=blue]{hyperref}

\usepackage{tabu}

\theoremstyle{plain}
\newtheorem{theorem}{Theorem}[section]
\newtheorem{lemma}{Lemma}[section]
\newtheorem{corollary}{Corollary}[section]

\newtheorem{example}{Example}[section]
\newtheorem{remark}{Remark}[section]
\newtheorem{assumption}{Assumption}

\usepackage{accents}

\renewcommand{\hat}{\widehat}
\renewcommand{\tilde}{\widetilde}
\renewcommand{\check}{\widecheck}

\newcommand{\ep}{\mathbb{E}}
\newcommand{\pr}{\mathbb{P}}
\newcommand{\var}{\mathrm{Var}}

\newcommand{\ii}{{\mathbbm{i}}}
\newcommand{\td}{{\mathrm{d}}}

\newcommand{\NEWnetworkerror}{\tilde {\cal M}_\alpha(\rho_n,n,\alpha;R)}

\usepackage{xcolor}
\definecolor{yuancolor4}{rgb}{0.56, 0.0, 1}
\definecolor{yuanyellow}{rgb}{1.0, 0.65, 0.0}
    \definecolor{yuancolor2}{rgb}{0.0, 0.4, 1.0}
    \definecolor{yuancolor3}{rgb}{0.0, 0.5, 0.0}
    \definecolor{yuancolor-edit}{rgb}{0.29, 0.59, 0.82}

\definecolor{yuan-british-green}{rgb}{0.0, 0.26, 0.15}

\definecolor{shaocolor}{RGB}{153,50,204}
\definecolor{mccolor}{rgb}{0.3010, 0.7450, 0.9330}
\definecolor{edgecolor}{rgb}{0, 0.5, 0}
\definecolor{respcolor}{rgb}{0.4940, 0.1840, 0.5560}
\definecolor{subscolor}{rgb}{0.8500, 0.3250, 0.0980}

\usepackage{hyperref}
\hypersetup{
    colorlinks=true,
    citecolor=blue,
    citebordercolor=red
}

\newcommand{\tOp}{{\tilde{O}_p}}

\newcommand{\da}{\vee}
\newcommand{\xiao}{\wedge}

\newcommand{\Mod}[1]{\ (\mathrm{mod}\ #1)}

\newcommand{\ma}{{M_\alpha}}

\newcommand{\jna}{{\cal J}_{n,\alpha}}

\newcommand{\ank}[1]{a_{n,\alpha;#1}}

\begin{document}

\def\spacingset#1{\renewcommand{\baselinestretch}%
{#1}\small\normalsize} \spacingset{1}

\if0\blind
{
    \title{\bf
    U-Statistic Reduction: Higher-Order Accurate Risk Control and Statistical-Computational Trade-Off, with Application to Network Method-of-Moments
    }
  \author{Meijia Shao\hspace{.2cm}\\
    Department of Statistics\\The Ohio State University\\
    and \\
    Dong Xia \\
    Department of Mathematics\\The Hong Kong University of Science and Technology\\
    and \\
    Yuan Zhang\\
    Department of Statistics\\The Ohio State University
    }
    \date{}
  \maketitle
} \fi

\if1\blind
{
  \bigskip
  \bigskip
  \bigskip
  \begin{center}
    {\LARGE\bf U-Statistic Reduction: Higher-Order Accurate Risk Control and Statistical-Computational Trade-Off, with Application to Network Method-of-Moments}
\end{center}
  \medskip
} \fi

\bigskip

\begin{abstract}
U-statistics play central roles in many statistical learning tools but face the haunting issue of scalability.
Significant efforts have been devoted into accelerating computation by U-statistic reduction.
However, existing results almost exclusively focus on power analysis, while little work addresses risk control accuracy -- comparatively, the latter requires distinct and much more challenging techniques.
In this paper, 
we establish the first statistical inference procedure with provably higher-order accurate risk control for incomplete U-statistics.
The sharpness of our new result enables us to reveal how risk control accuracy also trades off with speed for the first time in literature,
which complements the well-known variance-speed trade-off.
Our proposed general framework converts the long-standing challenge of formulating accurate statistical inference procedures for many different designs into a surprisingly routine task.
This paper covers non-degenerate and degenerate U-statistics, and network moments. 
We conducted comprehensive numerical studies and observed results that validate our theory's sharpness.
Our method also demonstrates effectiveness on real-world data applications.

\end{abstract}

\noindent%

\spacingset{1.5}

\section{Introduction}
\label{section::introduction}

\subsection{Background: U-statistics in statistical learning}

A \emph{U-statistic}, 
associated with $X_1,\ldots,X_n$ randomly sampled from a general probability space and a degree-$r$ permutation-invariant kernel function $h(x_1,\ldots,x_r)$ such that $h(x_1,\ldots,x_r) = h(x_{\pi(1)},\ldots,x_{\pi(r)})$ for any bijection $\pi:[1:r]\leftrightarrow[1:r]$, is defined as
\begin{equation}
    U_n
    :=
    \binom{n}r^{-1} \sum_{1\leq i_1<\ldots<i_r\leq n} h(X_{i_1},\ldots,X_{i_r})
    =:
    \binom{n}r^{-1} \sum_{I_r\in {\cal C}_n^r} h(X_{I_r}),
    \label{def::one-sample::complete-U-statistics}
\end{equation}
where ${\cal C}_n^k:=\{(i_1,\ldots,i_k): 1\leq i_1<\ldots<i_k\leq n\}$ and $X_{I_k}:=(X_{i_1},\ldots,X_{i_k})$ for $k\in[1:r]$.
U-statistics play central roles in many contemporary statistical learning methods.
\begin{example}[Example 1 of \citet{kong2020design}]
    \label{intro::example-1}
    Test the symmetry of the distribution of $X\in \mathbb{R}$ by
    $h(x_1,x_2,x_3):=
     {\rm sign}(2x_1-x_2-x_3)
    +{\rm sign}(2x_2-x_3-x_1)
    +{\rm sign}(2x_3-x_1-x_2)$.
\end{example}

\begin{example}[Bergsma-Dassios sign covariance \citep{bergsma2014consistent, moon2022interpoint}]
    \label{intro::example-2}
    To test the independence of $X\in\mathbb{S}_X$ and $Y\in\mathbb{S}_Y$, where $\mathbb{S}_X$ and $\mathbb{S}_Y$ are Banach spaces equipped with metrics $\rho_X$ and $\rho_Y$,
    define
    $
    h\big((x_1,y_1),\ldots(x_4,y_4)\big):=
    s_X(x_{i_1},\ldots,x_{i_4})
    s_Y(y_{i_1},\ldots,y_{i_4})
    $,
    where $s_X(t_1,\ldots,t_4) := {\rm sign}\big\{
        \rho_X(t_1,t_2) + \rho_X(t_3,t_4) - \rho_X(t_1,t_3) - \rho_X(t_2,t_4)
    \big\}
    $ and similarly define $s_Y$.
\end{example}

\begin{example}[Treatment effect measurement
\citep{rosenbaum2011new, zhao2018sensitivity}]
    \label{intro::example-3}
    Let $Y_1,\ldots,Y_n$ denote the observed treated-minus-control matched pair differences.  Given integers $1\leq \underline{r} \leq \overline{r}\leq r$, consider any $r$ observations $Y_{I_r}:=(Y_{i_1},\ldots,Y_{i_r})$, sorted by their absolute values: $|Y_{I_r(1)}|\leq \ldots\leq |Y_{I_r(r)}|$.  Define $h(Y_{I_r}):=\sum_{\ell=\underline{r}}^{\overline{r}} \mathbbm{1}_{[Y_{I_r(\ell)}>0]}$. Here $I_r(\ell)$ denotes the $\ell$-th element of $I_r$. 
\end{example}

\subsection{Computational challenge: why we need U-statistic reduction}

One major obstacle in the practical use of U-statistics lies in its heavy computational cost.  It takes $O(n^r)$ time even to evaluate the point estimator $U_n$.
For example, $r=2$ for Maximum Mean Descrepancy (MMD, assuming equal sample sizes) \citet{gretton2012kernel, schrab2021mmd}
and energy distance \citet{szekely2007measuring}; 
$r=3$ for testing distribution symmetry (Example \ref{intro::example-1});
$r=4$ for distance covariance (dCov) \citep{szekely2007measuring, yao2018testing} and sign covariance (Example \ref{intro::example-2}); and the new U-statistic in \citet{rosenbaum2011new} has a user-selected degree that can go up to 20-ish (Tables 3 and 4 in \citet{zhao2018sensitivity}).
Efforts to alleviate the computational challenge can be summarized into two mainstream directions.  The first direction explores shortcuts to fast-compute the complete U-statistic $U_n$.  For example, \citet{rosenbaum2011new} shows that its proposed U-statistic can be evaluated in $O(n\log n)$ time when the treatment effects are continuously distributed (no ties); also, \citet{huo2016fast} and \citet{chaudhuri2019fast} show that dCov can be computed in $O(n\log n)$ time when $X_i\in \mathbb{R}$;  
a similar acceleration is available for sign covariance with scalar inputs \citep{even2021counting}.  Unfortunately, most such shortcuts require scalar input.  For example, the Bergsma-Dassios sign covariance (Example \ref{intro::example-2}) with manifold-valued functional trajectories as input (see \citet{moon2022interpoint}, Simulation II) cannot be accelerated by \citet{heller2016computing} or \citet{even2021counting}
\footnote{To see why, notice that the key step in \citet{even2021counting} discusses the joint order relationships between the coordinates of $(x_1,y_1)$ and $(x_2,y_2)$.  This enables stripping the absolute value operations in $s_X(x_{[1:4]})s_Y(y_{[1:4]})$, where $\rho_X(u,v)=\rho_Y(u,v):=|u-v|$.  This trick is clearly inapplicable to non-scalar inputs.}.
Moreover, with non-scalar $X_i$ inputs, it may even be quite expensive to evaluate just one $h(X_{I_r})$ term.
For example, \citet{chakraborty2021new} computed the dissimilarity between two observed time series of earthquake waves \citep{UCRArchive2018} -- they directly computed an $\ell_2$ pairwise distance, but we may want to synchronize them before computing the distance \citep{srivastava2011registration, strait2019automatic}.  

The above discussion naturally leads us to the second acceleration strategy, namely, \emph{U-statistic reduction}, by averaging over a much smaller set of $r$-tuples in \eqref{def::one-sample::complete-U-statistics}.
Let 
\begin{align}
    \jna:=\big( I_r^{(1)},\ldots,I_r^{(|\jna|)} \big)
    \label{def::jna}
\end{align}
be a sequence of elements in ${\cal C}_n^r$ satisfying $|\jna|\asymp n^\alpha$ for some $\alpha\in(1,r)$.  We shall treat $\jna$ almost like a subset of ${\cal C}_n^r$, except that $\jna$ allows duplications.
The \emph{reduced U-statistic}, also called \emph{incomplete U-statistic} \citep{blom1976some,chen2019randomized}, with design $\jna$, is
\begin{align}
    U_J
    :=&~
    |\jna|^{-1}\sum_{I_r\in\jna}h(X_{I_r}).
    \label{def::general-reduced-U-stat}
\end{align}

\subsection{Two aspects of computational-statistical trade-off}
\label{subsec::intro::two-aspects-of-trade-off}

There are two kinds of prices we must pay for computation reduction.
First, this reduction inflates $\var(U_J)$, which further determines: (i) the confidence interval radius; and (ii) the minimum separation condition $|\mu_{H_a}-\mu_{H_0}|$ for consistently testing $H_0: \mu=\mu_{H_0}$ versus $H_a: \mu=\mu_{H_a}$. Here $\mu:=\ep U_n$. 
This aspect of computational-statistical trade-off is easy to quantify thus well-understood.
The overwhelming majority of existing literature on U-statistic reduction regards this aspect.
The pioneering work \citet{blom1976some} provides the earliest systematic study,
followed up by many works aiming at designing $\jna$ smartly to minimize $\var(U_J)$ under a given computational budget $O(n^{\alpha})$ \citep{lee1979asymptotic, lee1982incomplete, lee2019u, rempala2003incomplete,clemenccon2016scaling, kong2020design, durre2021consistency}.

The second kind of price for speeding-up is the depreciation of \emph{risk control accuracy} in statistical inference. The risk control accuracy refers to: (i) $|\pr(\textrm{true }\mu\in \textrm{CI}) - (1-\beta)|$ (confidence interval); and (ii) $|\pr(\textrm{actual type I error rate}) - \beta|$ (hypothesis testing), where the target confidence and significance levels are $1-\beta$ and $\beta$, respectively.
Technically speaking, it is governed by the accuracy of approximating the distribution of a properly-designed studentization of $U_J$.
Most existing works only prove asymptotic normality of $U_J$ \citep{brown1978reduced, janson1984asymptotic, chen2018distributed}; while to our best knowledge, \citet{chen2019randomized} is the only existing work that provides finite-sample error bound result in this regard.  
However, revealing the computational-statistical trade-off in risk control accuracy requires more accurate characterization of the sampling distribution of studentized incomplete U-statistic that is not yet available in existing literature\footnote{For example, the ``computational-statistical trade-off'' in Remark 3.1 of \citet{chen2019randomized} refers to the $\var(U_J)$ aspect; while their Berry-Esseen bound on distribution approximation (Theorem 3.3) does not characterize how risk control accuracy depreciates as computation reduces in the most interesting regime $N\gg n$.}.  Our paper aims to fill this blank.

\subsection{Our contributions}

In this article, we present the first comprehensive study on \emph{risk control accuracy} in statistical inference for incomplete U-statistics.
Our results cover non-degenerate, degenerate and noisy U-statistics.
For non-degenerate U-statistics, we established the first higher-order accurate distribution approximation to incomplete U-statistics with general designs.  
This leads to Cornish-Fisher CI's and tests with higher-order accurate risk controls.
Our approach only requires the design to satisfy two natural, weak and easy-to-verify assumptions.  
This converts the challenging task of analyzing many different designs into routine calculations.
Our method strictly complies with the $O(n^\alpha)$ computational budget and permits convenient parallelization.
For degenerate U-statistics, \citet{weber1981incomplete} and \citet{chen2019randomized} observed a very interesting phenomenon that the incompleteness of reduced U-statistics, apart from speeding up, also reinstates normality.  Pushing their results forward, we established the first finite-sample higher-order accurate distribution approximation for the studentization of an incomplete degenerate U-statistic.
Thirdly, we applied our theory and methods to incomplete network U-statistics, commonly known as \emph{network moments} \citep{bickel2011method, zhang2020edgeworth}.

Across all three scenarios, our method's accuracy significantly improves over the best existing results.
The sharpness of our error bounds allows us to reveal the trade-off between speed and the risk control accuracy of incomplete U-statistics, for the first time in literature.
For practitioners, 
our method provides fast and easy-to-implement solutions with guidance on how much computation would be required if the user aims at a certain risk control accuracy goal under \emph{finite sample size}.
Our analysis discovers interesting differences between the three scenarios.
They are reflected in the variation of practical guidance.
For non-degenerate U-statistics, achieving higher-order risk control accuracy only requires $\alpha>1$; and we can reduce the computation cost from $O(n^r)$ to $O(n^2)$ almost ``for free'': it does not deteriorate risk control accuracy at all and only inflates $\var(U_J)$ imperceptibly.
In sharp contrast, for most network moments (except the \emph{edge} motif), higher-order accurate inference requires $\alpha>2$ even for densest networks;
in presence of network sparsity, the guidance will further vary accordingly.

The theory part of this paper is majorly different from complete U-statistic literature \citet{helmers1991edgeworth,maesono1997edgeworth, putter1998empirical} and features several innovations.
The incompleteness nature of our studied U-statistic not only introduces new and complicated leading terms, but moreover breaches the symmetry of remainder terms, making existing bounds inapplicable.
We addressed this using a different technique.
A second example is that the formulation of a succinct and weak Assumption \ref{New-assumption-2}, a highlighted contribution of this paper, was also distilled from our theoretical exploration.
In the proof of Lemma \ref{lemma::random-jna-examples::check-assumption-2}, a key supporting result accompanying Assumption \ref{New-assumption-2}, we handled the challenging dependency in some random sampling schemes.
On nonparametric network analysis, our Theorem \ref{theorem::main-theorem::network-normality} significantly expands the classical analysis of triangles and V-shapes under sparse Erd\"os-Renyi model in \citet{gao2017testinga} to all motifs under general sparse graph models.  
The proof features our novel martingale analysis.

In summary, what our paper caters to is not just one specific application or data structure, but the infrastructural question of \emph{risk control accuracy} in U-statistic reduction.
The comprehensive theory \& methods toolbox we developed here provides an urgently needed (but currently missing) service to many U-statistic-based learning methods who want to properly control their inference risks while scaling up.

\section{Reduction of non-degenerate noiseless U-statistics}
\label{section::our-method}

\subsection{Method and theory for general designs}
\label{subsec::one-sample::general}

Recall that $\jna\subseteq {\cal C}_n^r$ denotes the index set for the reduced U-statistic $U_J$ defined in \eqref{def::general-reduced-U-stat}.
Following the convention \citep{lee2019u,chen2019randomized}, we call $\jna$ the \emph{design}.
Throughout this subsection, $\jna$ is given and fixed.
For any $I_k\in {\cal C}_n^k$, 
define $\ank{k}(I_k):=\big|\{ \tilde I_r\in \jna: I_k\subseteq \tilde I_r \}\big|$ to be the number of elements in $\jna$ that contain $I_k$ as a subset. An immediate fact is $\sum_{I_k\in {\cal C}_n^k} a_{n,\alpha;k}(I_k)={r\choose k} |\jna|$ for any $k\in[1:r]$. 
By \citet{han2018inference}, we have
\begin{align}
    U_J
    =&~
    \mu + |\jna|^{-1}\sum_{i=1}^n \ank{1}(i) g_1(X_i)
    +
    |\jna|^{-1}\sum_{k=2}^r\sum_{I_k\in {\cal C}_n^k} a_{n,\alpha;k}(I_k) g_k(X_{I_k}),
    \label{decomp::noiseless-U-stat-Hoeffding}
\end{align}
where $\mu:=\ep[h(X_{[1:r]})]$, and define projection terms:
$g_1(X_1):=\ep[h(X_{[1:r]})|X_1]-\mu$, for all $k\in [2:r]$,
$g_k(X_{[1:k]}):=\ep[h(X_{[1:r]})|X_{[1:k]}]-\mu-\sum_{k'=1}^{k-1}\sum_{I_{k'}\in {\cal C}_{[1:k]}^{k'}} g_{k'}(X_{I_{k'}})$. 
All $g_k$ terms are mutually uncorrelated. 
We call the U-statistic \emph{non-degenerate} if $\xi_1^2:=\var(g_1(X_1))\geq {\rm constant}>0$ and call it \emph{degenerate} if $\xi_1=0$.  
Here, we focus on the non-degenerate case and relegate the degenerate case to Section \ref{sec::one-sample::degenerate}.
In both Sections \ref{section::our-method} and \ref{sec::one-sample::degenerate}, we observe $h(X_{I_r})$ without observational error -- we call such U-statistics \emph{noiseless}.  In Section \ref{section::network-U-stat}, we study network moments as \emph{noisy} U-statistics, where we only observe noise-contaminated $h(X_{I_r})$'s.

Next, we estimate $\var(U_J)$ for studentizing $U_J$.  
This seemingly straightforward task \citep{han2018inference} now requires careful thoughts due to incompleteness.
Specifically, we need to choose between estimating $\sigma_J^2:=\var(U_J)$ or 
$\sigma_{J;1}^2:=\var(|\jna|^{-1}\sum_{i=1}^n \ank{1}(i) g_1(X_i))$.
This question does not exist in the complete U-statistic setting where the two quantities differ by $O(n^{-2})$ \citep{maesono1997edgeworth, zhang2020edgeworth}
, but now, we have
$|\sigma_J^2 - \sigma_{J;1}^2| \asymp n^{-\alpha}$,
which cannot be directly ignored when $\alpha\in(1,2]$.
In this paper, we choose to estimate
$\sigma_{J;1}$
for two reasons: formulation cleanness, and computation convenience.
Fortunately, the bias caused by using $\sigma_{J;1}$ instead of $\sigma_J$ to studentize $U_J$ will be properly accounted for by the Edgeworth correction terms, see Remark \ref{remark::where-is-trade-off-in-variance-reflected?}.
Now since $\sigma_{J;1}^2 = |\jna|^{-2} \big\{\sum_{i=1}^n \ank1^2(i)\big\} \xi_1^2$, it suffices to estimate $\xi_1^2 := \var(g_1(X_1))$ by
\begin{align}
    \tilde\xi_1^2
    :=&~
    n^{-\alpha} \sum_{i=1}^n\sum_{d=1}^{n^{\alpha-1}}
    h(X_{[i:d:(i+(r-1)d)]})h(X_{[i:(-d):(i-(r-1)d)]})
    -
    \tilde\mu^2,
    \label{def::xi_1-hat}
\end{align}
where 
$
    \tilde \mu^2
    :=
    n^{-\alpha} \sum_{i=1}^n\sum_{d=1}^{n^{\alpha-1}}
    h(X_{[i:d:(i+(r-1)d)]})
    h(X_{[(i+rd):d:(i+(2r-1)d)]}).
$
Computing $\tilde \xi_1^2$ costs $O(n^{\alpha})$. Denote the shorthand $M_{\alpha}:=|\jna|^{-1}\sum_{i=1}^n a_{n,\alpha;1}^2(i)\asymp n^{\alpha-1} $. Let $\{B_n\}_{n\geq 1}$ and $\{b_n\}_{n\geq 1}$ be sequences of random variables and deterministic values, respectively. We write $B_n=\tOp(b_n)$ if $\pr(B_n\geq C\cdot b_n)=O(n^{-1})$ for some absolute constant $C>0$ and when $n$ is large enough. The plug-in bias of $\tilde\xi_1^2$ is characterized by the following lemma. 

\begin{lemma}
    \label{lemma::variance-estimator}
    Set $\alpha\in (1,2)$, we have
    \begin{align}
        \tilde\xi_1^2 - \xi_1^2
        =&~
        \frac1n
        \sum_{i=1}^n\big\{
            g_1^2(X_i) - \xi_1^2
        \big\}
        +
        \frac1{n\ma}\sum_{i=1}^n\sum_{d=1}^{\ma}\sum_{\ell=1}^{r-1}
        g_1(X_i)
        \Big\{
            g_2(X_i,X_{i+\ell d}) + g_2(X_i,X_{i-\ell d})
        \Big\}
        \notag\\
        &+
        \tOp(n^{-\alpha/2}\log^{1/2}n).
        \label{lemma::xi-a-hat-decomp}
    \end{align}
\end{lemma}
With Lemma \ref{lemma::variance-estimator}, we are ready to formulate and decompose the studentization of $U_J$.
\begin{align}
    T_J
    :=&~
    \frac
    {U_J-\mu}
    {|\jna|^{-1}\big\{ \sum_{i=1}^n \ank{1}^2(i) \big\}^{1/2}\cdot \tilde\xi_1}
    =
    ({\cal T}_1 + {\cal T}_2)(1+ {\cal T}_3)^{-1/2}
    \notag\\
    =&~
    {\cal T}_1 + {\cal T}_2 -\frac12 
    {\cal T}_1
    {\cal T}_3
    +
    \tOp\big(n^{-\alpha/2}\log n\big),
\end{align}
where we define the following shorthand
\begin{align}
    {\cal T}_1
    :=&~
    \frac{\sum_{i=1}^n \ank1(i)g_1(X_i)}{\big\{\sum_{i=1}^n \ank1^2(i)\big\}^{1/2}\xi_1},
    \quad
    {\cal T}_2
    :=
    \frac{\sum_{k=2}^r \sum_{I_k\in{\cal C}_n^k}\ank{k}(I_k)g_k(X_{I_k})}{\big\{\sum_{i=1}^n \ank1^2(i)\big\}^{1/2}\xi_1},
    \notag\\
    {\cal T}_3
    :=&~
    \sum_{i=1}^n
    \frac{g_1^2(X_i) - \xi_1^2}{n\xi_1^2}
    +
    \frac1{n\ma \xi_1^2}\sum_{i=1}^n\sum_{d=1}^{\ma}\sum_{\ell=1}^{r-1}
    g_1(X_i)
    \Big\{
        g_2(X_i,X_{i+\ell d}) + g_2(X_i,X_{i-\ell d})
    \Big\}.
\end{align}

Next, we formulate the sampling distribution of $T_J$ with a fixed design $\jna$.  We aim at addressing general $\jna$'s, but of course it cannot be completely arbitrary.
Fortunately, it turns out that we only need two weak and natural regularity conditions on $\jna$.  

\begin{assumption}
    \label{New-assumption-1}
    The design of $\jna$ is data-oblivious, namely,
    \begin{align}
        \jna \perp (X_1,\ldots,X_n).
        \label{eqn::New-assumption-1}
    \end{align}
    For a deterministic $\jna$, \eqref{eqn::New-assumption-1} means $\jna$ is designed without consulting the data $X_{[1:n]}$.
\end{assumption}

Assumption \ref{New-assumption-1} is a very common assumption in U-statistic literature.  
We will briefly discuss a recent but seminal paper \citet{kong2020design} on data-aware designs in Section \ref{section::discussion}, but the quick gist is that being data-aware usually requires (possibly expensive) additional computation.
A second motivation for data-oblivious designs comes from the analysis of network U-statistics, where $X_{[1:n]}$ are latent \citep{zhang2020edgeworth}.

Our second assumption is motivated by the natural idea of avoiding bad designs such as $\jna:=\big([1:r],\ldots,[1:r]\big)$.
Existing literature suggests that a good design should be balanced, i.e., $g_k(X_{I_k})$'s with different $I_k$'s should have similar frequencies in \eqref{decomp::noiseless-U-stat-Hoeffding}.  Our next assumption formalizes this intuition.
\begin{assumption}
    \label{New-assumption-2}
    Set $\alpha\in (1,r)$.
    It holds simultaneously for all $k\in[1:r]$ and $I_k\in{\cal C}_n^k$ that
    \begin{align}
        \ank{k}(I_k)\in
        \begin{cases}
            [C_2,C_3]n^{\alpha-k}, & \textrm{ if }k<\alpha,
            \\
            [0,C_3]\log n, & \textrm{ if }k=\alpha,
            \\
            [0,C_3], & \textrm{ if }k>\alpha,
        \end{cases}
    \end{align}
    where $C_2,C_3: 0<C_2<C_3$ are universal constants.
    Moreover, for all $k\in[1:r]$
    \begin{align}
        \sum_{I_k\in {\cal C}_n^k} \ank k^2(I_k)
        \asymp&~
        n^\alpha\cdot n^{(\alpha-k)\da 0}.
    \end{align}
\end{assumption}
One might expect to see more assumptions, but surprisingly, Assumptions \ref{New-assumption-1} and \ref{New-assumption-2} are all we need.
This is one of the highlighted contributions of this paper.
These two assumptions are satisfied by our proposed deterministic ``no-waste'' design in Section \ref{subsec::our-reduction-design} and many popular randomized designs such as \citet{chen2019randomized} with high probability. We shall show in Section~\ref{subsec::analyasis-incomplete-u-stat} that four typical random designs satisfy Assumption~\ref{New-assumption-2} with high probability. 

Aside from Assumptions \ref{New-assumption-1} and \ref{New-assumption-2}, another common assumption in U-statistic literature is Cram\'er's condition \citep{helmers1991edgeworth, maesono1997edgeworth}.
\begin{equation}
    \limsup_{t\to\infty}\big|\ep[e^{\ii t \xi_1^{-1} \cdot g_1(X_1)}]\big|<1.
    \label{cramer-condition}
\end{equation}
The assumption \eqref{cramer-condition} is undesirably restrictive and violated by important applications, e.g., Example \ref{intro::example-1} with a discrete $X_1$ distribution.  To waive \eqref{cramer-condition}, inspired by \citet{lahiri1993bootstrapping} and \citet{shao2022higher}, we add to $T_J$ an \emph{artificial} smoothing term  $\delta_J\sim N(0,\sigma_\delta^2=C_{\delta_J}\log n\cdot n^{-\alpha/2})$ independent of $T_J$ with a large enough constant $C_{\delta_J}>0$.
We will show that $\delta_J$ waives \eqref{cramer-condition} by smoothing the CDF of $T_J$ without altering the distribution approximation formula and preserves the error bound\footnote{``Preserves the error bound'':  we can achieve the same error bound in approximating $F_{T_J+\delta_J}$, possibly without Cram\'er's condition, as that if we assume Cram\'er's condition and approximate $F_{T_J}$; where we use the same Edgeworth expansion formula across both cases.}.
Now we are ready to present our main result of this subsection.
Let
$\xi_k^2 := \var(g_k(X_{[1:k]}))$.
Define the population Edgeworth expansion formula for $T_J$ to be
\begin{align}
    &G_{\jna}(u)
    :=
    \Phi(u)
    +
    \phi(u)
    \cdot \Bigg[
        \Bigg(
            -\dfrac
                {\sum_{i=1}^n \ank1^3(i) (u^2-1)}
                {6\xi_1^3 \{\sum_{i'=1}^n \ank1^2(i')\}^{3/2}}
            +
            \dfrac
                {\sum_{i=1}^n \ank1(i) u^2}
                {2\{\sum_{i'=1}^n \ank1^2(i')\}^{1/2} n\xi_1^3}
        \Bigg)
        \ep[g_1^3(X_1)]
        \notag\\
        &+
        \Bigg(
            -\dfrac
                {\sum_{1\leq i<j\leq n} \ank1(i)\ank1(j)\ank2(\{i,j\}) (u^2-1)}
                {\{\sum_{i'=1}^n \ank1^2(i')\}^{3/2}\xi_1^3}
            \notag\\
            &+
            \dfrac
                {(r-1)\sum_{i=1}^n \ank1(i) u^2}
                {\{\sum_{i'=1}^n \ank1^2(i')\}^{1/2}n\xi_1^3}
        \Bigg)
        \ep[g_1(X_1)g_1(X_2)g_2(X_1,X_2)]
        -
        \sum_{\ell=1}^{\lfloor \frac{\alpha/2}{\alpha-1} \rfloor}
         \Bigg\{
            \frac{H_{2\ell-1}(u)}{(2\ell)!\{\sum_{i'=1}^n \ank1^2(i')\}^\ell\xi_1^{2\ell}}
            \notag\\
            &\times
            \sum_{k_1,\ldots,k_\ell\in[2:r]} 
                \sum_{\substack{
                    \big(I_{k_1}^{(1)},\ldots,I_{k_\ell}^{(\ell)}\big)\in {\cal C}_n^{k_1}\otimes \cdots \otimes {\cal C}_n^{k_\ell}\\
                    \textrm{$I_{k_1}^{(1)},\ldots,I_{k_\ell}^{(\ell)}$ mutually disjoint}
                }}
                \ank{k_1}^2(I_{k_1}^{(1)}) \cdots \ank{k_\ell}^2(I_{k_\ell}^{(\ell)})
                \xi_{k_1}^2\cdots \xi_{k_\ell}^2
        \Bigg\}
    \Bigg],
    \label{Edgeworth::generic}
\end{align}
where $H_k(u):=(-1)^k e^{u^2/2}\td^k/\td u^k (e^{-u^2/2})$ is the $k$th Hermite polynomial (\citet{slater1960confluent}, page 99).
In practice, we use the empirical version of \eqref{Edgeworth::generic} with estimated coefficient components.
We stress that under many $\jna$ designs, the last term in \eqref{Edgeworth::generic} can be greatly simplified, thus we still comply with the $O(n^\alpha)$ computation (see Sections \ref{subsec::our-reduction-design}, \ref{subsec::analyasis-incomplete-u-stat}).
Define
\begin{align}
    \tilde \ep[g_1^3(X_1)]
    :=
    \frac1n \sum_{i=1}^n h(X_{[i:(i+r-1)]})& h(X_{\{i,[(i+r):(i+2r-2)]\}}) h(X_{\{i,[(i+2r-1):(i+3r-3)]\}})
    -\tilde{\mu}^3
    \label{def::g_1^3-hat}
    \\
    \tilde \ep[g_1(X_1)g_1(X_2)g_2(X_1,X_2)]
    :=&~
    \frac1n \sum_{i=1}^n h(X_{[(i-r+1):i]}) h(X_{[i:(i+r-1)]}) h(X_{[(i+r-1):(i+2r-2)]})
    \notag\\
    &-\tilde{\mu}^3 - 2U_J \cdot \tilde\xi_1^2
    \label{def::g_1g_1g_2-hat}
    \\
    \tilde\xi_k^2
    :=
    \frac{1}{n^\alpha}\sum_{i=1}^n\sum_{d=1}^{n^{\alpha-1}} h(X_{[i:d:(i+(r-1)d)]})&h(X_{[(i+(k-1)d):(-d):(i-(r-k)d)]})-\tilde\mu^2-\sum_{k'=1}^{k-1}\binom{k}{k'} \tilde\xi_{k'}^2,
    \label{def::xi_k-hat}
\end{align}
for $k\in[2:r]$.
Let $\tilde G_{\jna}(u)$ be the empirical version of $G_{\jna}(u)$
with coefficients estimated by
\eqref{def::xi_1-hat}, \eqref{def::g_1^3-hat}, \eqref{def::g_1g_1g_2-hat} and \eqref{def::xi_k-hat}.
We will also write $G_J(u)$ in short for $G_{\jna}(u)$, especially in the proof. 
Now we present our main theorem on the distribution approximation error.
\begin{theorem}
    \label{theorem::general-main-theorem}
    Set $\alpha\in(1,2]$.  If $\xi_1>0$ and $\jna$ satisfies Assumptions \ref{New-assumption-1} and \ref{New-assumption-2}, then
    \begin{align}
        \big\|
            F_{T_J+\delta_J|\jna}(u) - G_{\jna}(u)
        \big\|_\infty
        =&~
        O\big(
            n^{-\alpha/2}\log^{1/2}n
        \big)
        \label{eqn::theorem::non-degenerate::general-1}
        \\
        \big\|
            F_{T_J+\delta_J|\jna}(u) - \tilde G_{\jna}(u)
        \big\|_\infty
        =&~
        \tOp(n^{-\alpha/2}\log^{1/2}n).
        \label{eqn::theorem::non-degenerate::general-2}
    \end{align}
    If $\jna$ is deterministic, then $F_{T_J+\delta_J|\jna}(u)$ should be understood as $F_{T_J+\delta_J}(u)$.
\end{theorem}
\begin{remark}
    \label{remark::noiseless-why-not-alpha-greater-than-2}
    Theorem \ref{theorem::general-main-theorem} provides the practical guidance that for non-degenerate U-statistics, paying more than $O(n^2)$ computation (i.e. ,$\alpha>2$) will not further merit risk control accuracy, since the error bound at $\alpha=2$ already matches the risk control accuracy of the complete U-statistic \citep{helmers1991edgeworth,maesono1997edgeworth}.  
    Also, increasing $\alpha$ beyond 2 only brings $O(n^{-2})$ improvement to $\var(U_J)$ \citep{lee2019u}.
    Considering the computational cost grows exponentially in $\alpha$, it is therefore not worthwhile to set $\alpha$ above 2 in this scenario.
\end{remark}
\begin{remark}
    Remark 3.1 in \citet{chen2019randomized} points out that as $\alpha$ decreases towards 1, then $\sigma_{J;1}$ becomes a poorer approximation to $\sigma_J$; when $\alpha=1$, their difference no longer diminishes with $n\to\infty$.
    This is called \emph{phase change} in \citet{weber1981incomplete, chen2019randomized}.
    While \citet{weber1981incomplete, chen2019randomized} exclusively focused on $\var(U_J)$, 
    we reveal an interesting second aspect of what happens as $\alpha\to 1$: the Edgeworth expansion becomes lengthier, and the risk control accuracy also depreciates.
    If we do not incorporate an increasing number of bias-correction terms in the Edgeworth expansion, it will depreciate the risk control accuracy even faster (the $n^{-\alpha/2}$ term in Theorem \ref{theorem::general-main-theorem} will be replaced by $n^{-(\alpha-1)}$, which is the Berry-Esseen bound of the normal approximation to $T_J$ that uses an estimated $\sigma_{J;1}$ to studentize).
\end{remark}

Next, we invert the Edgeworth expansion to formulate the Cornish-Fisher confidence interval with higher-order accurate confidence level control.
This inversion, however, is a quite nontrivial generalization of the convention \citep{hall1983inverting, hall2013bootstrap},
because the Edgeworth expansion of $T_J$ contains not only a $O(n^{-1/2})$ term, but also $O\big(n^{-(\alpha-1)\ell}\big)$, $\ell\in\{1,\ldots,\lfloor\alpha/(2(\alpha-1))\rfloor\}$ terms.
To ease notation, we summarize \eqref{Edgeworth::generic} into the following prototype formula
\begin{align}
    G_{\jna}(u)
    :=&~
    \Phi(u)
    +
    \phi(u)
    \Bigg\{
        \frac{\Gamma_0(u)}{\sqrt{n}}
        +
        \sum_{\ell=1}^{\lfloor \frac{\alpha/2}{\alpha-1} \rfloor}
        \frac{\Gamma_\ell(u)}{\ma^\ell}
    \Bigg\}
    \label{eqn::Edgeworth::prototype}
\end{align}
Recall that we set $\ma := |\jna|^{-1} \sum_{i=1}^n\ank1^2(i)\asymp n^{\alpha-1}$ due to Assumption \ref{New-assumption-2}.
The other components $\Gamma_0(\cdot)$ and $\Gamma_\ell(\cdot)$'s can be easily defined by comparing \eqref{Edgeworth::generic} to \eqref{eqn::Edgeworth::prototype}.
Given any user-selected significance level $\beta\in (0,1/2)$, define the Cornish-Fisher expansion terms $\Psi_0(z_\beta)$ and $\Psi_\ell(z_\beta)$ for $\ell=1,\ldots,\lfloor(\alpha/2)/(\alpha-1)\rfloor$ recursively as follows.  First, set
\begin{align}
    \Psi_0(z_\beta)
    :=&~
    \Gamma_0(z_\beta),
    \quad
    \Psi_1(z_\beta)
    :=
    -\Gamma_1(z_\beta).
    \label{method::one-sample::CF-eqn-0-1}
\end{align}
Then for each $k\in\big[2:\lfloor(\alpha/2)/(\alpha-1)\rfloor\big]$ and with the already-defined $\Psi_0(z_\beta),\ldots,\Psi_{k-1}(z_\beta)$, we define $\Psi_k(z_\beta)$ by solving the following recursive equation
\begin{align}
    &\Psi_k(z_\beta)
    =
    -\sum_{\ell'=2}^k
    \Bigg\{
        \sum_{\substack{
            j_1,\ldots,j_{\ell'}:\\
            1\leq \{j_1,\ldots,j_{\ell'}\} \leq k-\ell'+1\\
            j_1+\cdots+j_{\ell'}=k
        }}
        \Psi_{j_1}(z_\beta)\cdots \Psi_{j_{\ell'}}(z_\beta)
        \cdot \frac{\phi^{(\ell'-1)}(z_\beta)}{(\ell')!}
    \Bigg\}
    \notag\\
    &
    -
    \sum_{\substack{
        k_1,k_2: k_1+k_2=k\\
        k_1=0,\ldots,k-1\\
    }}
    \Bigg[
        \Bigg\{
            \sum_{\ell'=1}^{k_1}\sum_{\substack{
                j_1,\ldots,j_{\ell'}:\\
                1\leq \{j_1,\ldots,j_{\ell'}\} \leq k-\ell'+1\\
                j_1+\cdots+j_{\ell'}=k
            }}
            \Psi_{j_1}(z_\beta)\cdots \Psi_{j_{\ell'}}(z_\beta)
            \cdot \frac{\phi^{(\ell'-1)}(z_\beta)}{(\ell')!}
        \Bigg\}
        \notag\\
        &
        \cdot 
        \Bigg\{
            \Gamma_{k_2}(z_\beta)
            +
            \sum_{\ell'=1}^{k_2-1}
            \sum_{\ell''=1}^{k-\ell'}
            \Bigg\{
                \sum_{\substack{
                    j_1,\ldots,j_{\ell''}:\\
                    1\leq \{j_1,\ldots,j_{\ell''}\} \leq k-\ell'-\ell''+1\\
                    j_1+\cdots+j_{\ell''} = k-\ell'
                }}
                \Psi_{j_1}(z_\beta)\cdots \Psi_{j_{\ell''}}(z_\beta)
                \cdot
                \frac{\Gamma_{\ell'}^{(\ell'')}(z_\beta)}{(\ell'')!}
            \Bigg\}
        \Bigg\}
    \Bigg].
    \label{method::one-sample::CF-eqn-2-induction}
\end{align}
Our population Cornish-Fisher expansion that approximates the lower $\beta$-quantile of $T_J$ is
\begin{align}
    G_{\jna}^{-1}(z_\beta)
    =:&~
    z_\beta
    -
    \frac{\Psi_0(z_\beta)}{\sqrt n}
    +
    \sum_{\ell=1}^{\lfloor \frac{\alpha/2}{\alpha-1} \rfloor} \frac{\Psi_\ell(z_\beta)}{\ma^\ell}.
    \label{def::one-sample::non-degenerate::C-F}
\end{align}
Next, to define the empirical C-F expansion, we first re-express the empirical Edgeworth expansion $\tilde G_{\jna}$ in the format of \eqref{eqn::Edgeworth::prototype}, which defines
the empirical versions of $\Gamma_0,\Gamma_1,\ldots$, denoted by $\tilde \Gamma_0, \tilde \Gamma_1,\ldots$.
Replacing $\Gamma$'s in \eqref{method::one-sample::CF-eqn-0-1} and \eqref{method::one-sample::CF-eqn-2-induction} by $\tilde \Gamma$'s, we obtain the empirical $\tilde\Psi$'s
and thus the empirical C-F expansion
$\tilde G_{\jna}^{-1}(\cdot)$.
We omit further details due to page limit.
\begin{theorem}
    \label{theorem::one-sample::non-degenerate::main-theorem-CF-expansion}
    Under the conditions of Theorem \ref{theorem::general-main-theorem}, for any given $\beta\in(0,1)$,
    the Cornish-Fisher expansion \eqref{def::one-sample::non-degenerate::C-F} satisfies
    \begin{align}
        F_{T_J+\delta_J|\jna}\big( G_{\jna}^{-1}(z_\beta) \big)
        =&~
        \beta + O\big(n^{-\alpha/2}\big)
        \\
        F_{T_J+\delta_J|\jna}\big( \tilde G_{\jna}^{-1}(z_\beta) \big)
        =&~
        \beta + \tOp\big(n^{-\alpha/2}\log^{1/2}n\big).
    \end{align}
\end{theorem}
Theorems \ref{theorem::general-main-theorem} and \ref{theorem::one-sample::non-degenerate::main-theorem-CF-expansion} enable inference methods with higher-order accurate risk control.
\begin{corollary}
    \label{corollary::one-sample::inference}
    Under the conditions of Theorem \ref{theorem::general-main-theorem}, we have
    \begin{enumerate}[(a).]
        \item \label{corollary::one-sample::inference::part-a}
        Test the hypotheses
        \begin{equation}
            H_0: \mu = \mu_0;
            \quad \textrm{vs.}\quad
            H_a: \mu \neq \mu_0
            \notag\\
        \end{equation}
        using the following Empirical p-value, denoted by $\mathfrak{p}$
        \begin{align}
            \mathfrak{p}
            :=&~
            2\min\Big\{
                \tilde G_{\jna}( T_J^{(\rm obs)}+\delta_J),
                1-\tilde G_{\jna}( T_J^{(\rm obs)}+\delta_J)
            \Big\},
            \label{non-degenerate::p-value}
        \end{align}
        where $T_J^{(\rm obs)}:=(U_J-\mu_0) / \big\{|\jna|^{-1}\{\sum_{i=1}^n \ank1^2(i)\}^{1/2} \tilde\xi_1\big \}$.
        This test enjoys higher-order accurate type-I error control: 
        \begin{align}
            \pr_{H_0}\big(
                \mathfrak{p}<\beta
                \big|
                \jna
            \big)
            =&~
            \beta + O(n^{-\alpha/2}\log^{1/2}n).
            \notag
        \end{align}
        
        \item \label{corollary::one-sample::inference::part-b}
        The Cornish-Fisher confidence interval ${\cal I}_\beta$ defined by 
        \begin{align}
            {\cal I}_\beta
            :=
            \Big(
                &U_J - (\tilde G_{\jna}^{-1}(z_{1-\beta/2})-\delta_J)
                    \cdot
                    |\jna|^{-1}\big\{ \sum_{i=1}^n \ank{1}^2(i) \big\}^{1/2}\cdot \tilde\xi_1
                ,
                \notag\\
                &
                U_J - (\tilde G_{\jna}^{-1}(z_{\beta/2})-\delta_J)
                    \cdot
                    |\jna|^{-1}\big\{ \sum_{i=1}^n \ank{1}^2(i) \big\}^{1/2}\cdot \tilde\xi_1
            \Big)
            \notag
        \end{align}
        enjoys a higher-order accurate control of actual coverage probability around $1-\beta$:
        \begin{align}
            \pr\big(
                \mu
                \in
                {\cal I}_\beta
                \big|
                \jna
            \big)
            =&~
            1 - \beta + O(n^{-\alpha/2}\log^{1/2}n).
            \label{def::general-Cornish-Fisher-CI}
        \end{align}
    \end{enumerate}
    In both parts, $\pr(\cdot |\jna)$ are taken with respect to the randomness of both $X_{[1:n]}$ and $\delta_J$.
\end{corollary}

As mentioned in Section \ref{subsec::intro::two-aspects-of-trade-off}, reducing the U-statistic inflates $\var(U_J)$.
However, we studentize $U_J$ by $\tilde\sigma_{J;1}$, which only captures the leading term in $\var(U_J)$, whose order does \emph{not} vary with $\alpha$.  
Readers naturally wonder where the variance inflation is reflected in our statistical inference procedure.
Here, we use our CI formula as an example to clarify. 
\begin{remark}
    \label{remark::where-is-trade-off-in-variance-reflected?}
    The radius of our Cornish-Fisher CI is $O\big(n^{-1/2}+n^{-(\alpha-1/2)}\big)$\footnote{
        To see this, notice that $\Gamma_0(-u)=\Gamma_0(u)$, while $\Gamma_\ell(-u)=-\Gamma_\ell(u)$ for all $\ell\geq 1$.  
        Also notice that 
        $(\tilde G_{\jna}^{-1}(z_{1-\beta/2})-\delta_J)\asymp 1+n^{-(\alpha-1)}$ and $|\jna|^{-1}\big\{ \sum_{i=1}^n \ank{1}^2(i) \big\}^{1/2}\asymp n^{-1/2}$.
    }.
    Studentizing $U_J$ with $\tilde \sigma_J$ will also yield a CI radius of $\{O(n^{-1}+n^{-\alpha})\}^{1/2} = O\big(n^{-1/2}+n^{-(\alpha-1/2)}\big)$.  
    In other words, using $\tilde\sigma_J$ or $\tilde\sigma_{J;1}$ to studentize $U_J$ lead to different pivots as intermediate steps, but eventually, they will produce essentially the same Cornish-Fisher CI's.
\end{remark}
We conclude this subsection with the observation that any test, based on an asymptotically normal $T_J$, that correctly estimates the leading term of $\var(U_J)$, namely, $\sigma_J^2$, is asymptotically power-optimal (see how Theorem 3.5 of \citet{banerjee2017optimal} establishes the asymptotic power-optimality of their test).  This of course also applies to our method.
We reiterate that power-optimality is conceptually distinct from risk control accuracy (see Section \ref{subsec::intro::two-aspects-of-trade-off}).

\subsection{Applications: statistical inference with higher-order accurate risk control for deterministic and randomized designs}
\label{subsec::non-degenerate::applications}

In this subsection, we apply our general framework in Section \ref{subsec::one-sample::general} to analyzing several popular designs.  In Section \ref{subsec::our-reduction-design}, we propose and analyze a novel variance-optimal deterministic reduction scheme.
In Section \ref{subsec::analyasis-incomplete-u-stat}, we present the first provably higher-order accurate inference for randomized incomplete U-statistics \citep{lee2019u,chen2019randomized}.

\subsubsection{A novel variance-optimal deterministic design}
\label{subsec::our-reduction-design}

As aforementioned, most past works on U-statistic reduction focus on designing a deterministic $\jna$ that minimizes $\var(U_J)$.  
This task demands a design $\jna$, under which all $\ank{k}(I_k)$'s take as similar values as possible; specifically, for $k>\alpha$, we desire $\ank{k}(I_k)\in\{0,1\}$.
Despite the significant efforts in existing literature, it remains a challenging open problem to construct such $\jna$ for a general $\alpha>1$.  
However, our analysis reveals that increasing $\alpha$ above 2 brings little merit to both inference power and risk control accuracy.  Therefore, we can naturally consider the much more doable problem with $\alpha\in(1,2]$.  
Slightly abusing notation, set $\ma:=\lfloor n^{\alpha-1} \rfloor$\footnote{This $\ma$ will later play the same role as the $\ma$ defined in \eqref{def::one-sample::non-degenerate::C-F}, so we do not set up two symbols.}.
Define an $r$-tuple ${\cal A}_{i,d}^{n,r}$ as follows.
\begin{align}
    {\cal A}_{i,d}^{n,r}
    := &
    \Big\{
        (i_1,\ldots,i_r):
        1\leq i_k\leq n,
        i_k \equiv j_k \Mod{n},
        j_k = i+(2^{k-1}-1)d,
        \forall k\in[1:r]
    \Big\}.
    \label{def::A-index-set}
\end{align}
Our proposed deterministic design $\jna$ is
\begin{align}
    \jna
    := &
    \bigcup_{d=b_1 \ma}^{b_2 \ma} \bigcup_{i=1}^n {\cal A}_{i,d}^{n,r},
    \label{def::jma}
\end{align}
where $b_1,b_2:0<b_1<b_2$ are constants.
It is not difficult to check that $|\jna|=O(n^\alpha)$.
\begin{lemma}
    \label{lemma::our-method-no-waste-theorem}
    Suppose $n\gg r$.  Set $\alpha\in(1,2)$ and $b_1/b_2\in \big( (2^{r-1}-1)/(2^{r-1}), 1 \big)$.
    Our design \eqref{def::A-index-set} and \eqref{def::jma} 
    satisfies
    $\ank1(i) \asymp\ma$ for all $i\in [1:n]$ and $\ank{k}(I_k)\in\{0,1\}$, for all $k\geq 2$  and all $I_k\in{\cal C}_n^k$. Moreover, $\sum_{I_k\in {\cal C}_n^k}a_{n,\alpha;k}^2(I_k)\asymp n^{\alpha+(\alpha-k)\vee 0}$ for all $k\in[1:r]$. 
\end{lemma}

Lemma \ref{lemma::our-method-no-waste-theorem} implies that our design achieves the optimal variance among all data-oblivious reduction schemes under $O(n^\alpha)$ computational budget limit.
As an immediate corollary of Lemma \ref{lemma::our-method-no-waste-theorem}, our deterministic $\jna$ also satisfies Assumption \ref{New-assumption-2}, so Theorem \ref{theorem::general-main-theorem} and Corollary \ref{corollary::one-sample::inference} hold.
It has a handy Edgeworth expansion simplification
\begin{align}
    \Gamma_0(u)
    :=&~
    \frac{2u^2+1}{6\xi_1^3} \ep[g_1^3(X_1)]
    +
    \frac{(r-1)(u^2+1)}{2\xi_1^3} \ep[g_1(X_1)g_1(X_2)g_2(X_1,X_2)]
    \label{term::noiseless-U-stat-ma-terms-0}\\
    \Gamma_\ell(u)
    :=&~
    -\Bigg\{
        \frac{\sum_{k=2}^r \xi_k^2\binom{r}k}{(b_2-b_1)r^2\xi_1^2}
    \Bigg\}^\ell
    \cdot
    \frac{H_{2\ell-1}(u)}{(2\ell)!}
    =
    -\Bigg\{
        \frac{\sigma_h^2-r\xi_1^2}{(b_2-b_1)r^2\xi_1^2}
    \Bigg\}^\ell
    \cdot
    \frac{H_{2\ell-1}(u)}{(2\ell)!}
    \label{term::noiseless-U-stat-ma-terms}
\end{align}
for all $\ell=1,\ldots,\lfloor \alpha/\{2(\alpha-1)\} \rfloor$,
where $\sigma_h^2:=\var(h(X_{[1:r]}))$.
We can estimate $\sigma_h^2$ by
\begin{align}
    \tilde\sigma_h^2
    :=&~
    \frac{1}{n\ma} \sum_{i=1}^n\sum_{d=1}^\ma h^2(X_{[i:d:(i+(r-1)d)]}) - \tilde\mu^2,
    \label{est::our-design-sigma-h}
\end{align}
where $\tilde\mu$ is inherited from \eqref{def::xi_1-hat}.
Plugging \eqref{term::noiseless-U-stat-ma-terms-0}, \eqref{term::noiseless-U-stat-ma-terms} and our setting $\ma=\lfloor n^{\alpha-1}\rfloor$ into \eqref{eqn::Edgeworth::prototype}, we obtain the population Edgeworth expansion $G_J(u)$ for $T_J+\delta_J$ under our $\jna$.
Using the plug-in estimators \eqref{def::g_1^3-hat}, \eqref{def::g_1g_1g_2-hat} and \eqref{est::our-design-sigma-h}, we obtain the empirical Edgeworth expansion $\tilde G_J(u)$.  Finally, using \eqref{method::one-sample::CF-eqn-0-1},  \eqref{method::one-sample::CF-eqn-2-induction}, and \eqref{def::one-sample::non-degenerate::C-F}, we can formulate the Cornish-Fisher CI.
Theoretical guarantees for our proposed design are straight corollaries of our main theorems in Section \ref{subsec::one-sample::general} and thus omitted here due to page limit.

Interestingly, our method, while being an acceleration tool itself, can also service some other acceleration tools and make them even faster.
One such ``customer'' is the \emph{divide-and-conquer acceleration} by parallel computing in \citet{chen2018distributed}.
We have $K$ parallel computing servers who will return summary statistics to the main server for aggregation.
In \citet{chen2018distributed}, each individual server still computes a \emph{complete} U-statistic, thus readers already see that our method can significantly improve on this point.
Due to page limit, we sink the details of coupling our method with \citet{chen2018distributed} to Algorithm 1 in Supplementary Material.
We stress that Algorithm 1 can also be viewed as a parallel-computing strategy for implementing our inference method on several servers, while the inference formula and accompanying theory remain unchanged.
We compare our method coupled with \citet{chen2018distributed} versus the vanilla \citet{chen2018distributed} in Table \ref{tab::comparison-chen-peng-reduction}, where for cleanness, we equate all split sizes, set $K\asymp n^{\tau'}$ as in \citet{chen2018distributed} and equate the orders of the second leading terms in the two approaches' variances by setting $\alpha=2-\tau'$.
Table \ref{tab::comparison-chen-peng-reduction} shows that our method significantly accelerates \citet{chen2018distributed}, with little to no variance inflation (depending on $\tau'$ vs $\alpha$) and higher accuracy in risk control.

\begin{table}[h!]
    \centering
    \addtolength{\leftskip} {-1.5cm}
    \caption{Merits of our method coupled with \citet{chen2018distributed}.  Set $\alpha\in(1,2)$.}
    \label{tab::comparison-chen-peng-reduction}
    \begin{tabular}{ccc}\hline
         & Vanilla \citet{chen2018distributed} & Our method + \citet{chen2018distributed}
         \\\hline
         Time cost on each server & 
            $O(n^r/K^{r-1})$ & 
            $O\big((n/K)^{\alpha-1}\cdot K\big)$
         \\
         Variance of aggregated U-stat. &
            $r^2\xi_1^2/n + O(Kn^{-2})$ & $r^2\xi_1^2/n + O(n^{-\alpha})$
         \\
         CDF approximation error &
            $o(n^{-1/2})$\tablefootnote{This requires $K=O(n^{\tau'})$ for $\tau'\in(0,1/4)$.  See Theorem 3.3-(i) in \citet{chen2018distributed}.}
            & $O(n^{-\alpha/2})$
         \\
         Inference error\tablefootnote{Inference error may be represented by the discrepancy between the nominal and the actual CI coverage probabilities or that between the nominal significance level and the actual type I error rate.}
         &
            $o_p(1)$
            \tablefootnote{
                The vanilla \citet{chen2018distributed} \emph{standardizes} $U_J$.
                However, standardization-based inference needs a bias-correction that de-facto reproduces the result of the studentization-based inference in order to achieve higher-order accuracy.  See \citet{hall2013bootstrap}, Section 3.10.2.
            }
            & $\tOp(n^{-\alpha/2}\log^{1/2}n)$
         \\\hline
    \end{tabular}
\end{table}

\subsubsection{Analysis of randomized incomplete U-statistics}
\label{subsec::analyasis-incomplete-u-stat}

The general framework we propose in Section \ref{subsec::one-sample::general} is particularly convenient for analyzing randomized designs.
Here, we showcase the analysis for a few example designs in literature:
\begin{enumerate}[start=1, label={(J\arabic*)}]
    \item \label{jna-choice-1:SRS-w-replace}
    Sample $n^\alpha$ size-$r$ subsets from ${\cal C}_n^r$ at random, with replacement.
    
    \item \label{jna-choice-2:SRS-wo-replace}
    Similar to \ref{jna-choice-1:SRS-w-replace} but sample without replacement\footnote{Notice that sampling $\jna:|\jna|=O(n^\alpha)$ without replacement could also be done with $O(n^\alpha)$ budget, in both time and memory, by a lexicographic indexing of elements in ${\cal C}_n^r$.}.
    
    \item \label{jna-choice-3:stratified-w-replace}
    For $i=1,\ldots,n$, sample $n^{\alpha-1}$ size-$r$ subsets from ${\cal C}_n^r$ containing $i$, with replacement.
    
    \item \label{jna-choice-4:stratified-wo-replace}
    Similar to \ref{jna-choice-3:stratified-w-replace}, but for each $i$,  sample without replacement.
\end{enumerate}
These sampling schemes are natural to come up with, and there are many more similar randomized designs in existing literature.
However, no available theory and methods yet exist to address the problem of higher-order accurate risk control for any of them.
The conventional analysis typically starts with re-expressing $U_J$ as follows \citep{chen2019randomized}.
\begin{align}
    U_J-\mu
    :=&~
    \underbrace{(U_n-\mu)}_{\textrm{(Part I)}}
    +
    \underbrace{|\jna|^{-1}\sum_{I_r\in\jna}\big\{ h(X_{I_r})-U_n \big\}}_{\textrm{(Part II)}}
    =:
    (U_n-\mu) + V_J,
    \label{analysis::randomized-incomplete-U-stat-conventional}
\end{align}
where part I is a rescaled complete U-statistic (see definition in eq. (\ref{def::one-sample::complete-U-statistics})) and part II captures the randomness in $\jna$.
Then one can normal-approximate both parts, respectively, to establish the asymptotic normality of $U_J$ via careful conditioning and convolution, see pages 9--20 in \citet{chen2019randomized-SUPP}.
While \eqref{analysis::randomized-incomplete-U-stat-conventional} is very useful for analyzing degenerate U-statistics (see our Section \ref{sec::one-sample::degenerate}), it is not a sharp tool in the non-degenerate case, where both parts will contribute non-ignorable terms to the Edgeworth expansion of studentized $U_J$.
Therefore, our analysis here takes a very different route: the key is to apply our general framework in Section \ref{subsec::one-sample::general}.
To this end, we first verify that our considered randomized design $\jna$ satisfies Assumption \ref{New-assumption-2} with high probability.  (Assumption \ref{New-assumption-1} is easily verified.)

\begin{lemma}
    \label{lemma::random-jna-examples::check-assumption-2}
    Let $\jna$ be constructed by one of \ref{jna-choice-1:SRS-w-replace}--\ref{jna-choice-4:stratified-wo-replace}.
    For any given constant $C_1>0$, there exist constants $C_2,C_3: C_3>C_2>0$ depending on $C_1$ and the design $\jna$, such that Assumption \ref{New-assumption-2} holds with probability at least $1-n^{-C_1}$.
\end{lemma}
Next, we showcase the formula simplification for \ref{jna-choice-1:SRS-w-replace} and \ref{jna-choice-3:stratified-w-replace} by taking another layer of expectation with respect to the randomness of $\jna$, denoted by $\ep_J$, as follows.

\begin{corollary}
    \label{corollary::jna-J1-J3-simplification}
 Set $\alpha\in(1,2)$.  For randomized deign \ref{jna-choice-1:SRS-w-replace}, we have
    \begin{align}
        \ep_J[\Gamma_0(u)]
        :=&~
        \frac{2u^2+1}{6\xi_1^3}\ep[g_1^3(X_1)]
        +
        \frac{(r-1)(u^2+1)}{2\xi_1^3}\ep[g_1(X_1)g_1(X_2)g_2(X_1,X_2)]
        \label{def::random-incomplete-U-stat-G-check-0}
        \\
        \ep_J[\Gamma_\ell(u)]
        :=&~
        -
        \frac{H_{2\ell-1}(u)}{(2\ell)!}
        \Big\{
            \frac
                {\sum_{k=2}^r \binom rk\xi_k^2}
                {r^2\xi_1^2}
        \Big\}^\ell.
        \label{def::random-incomplete-U-stat-G-check-ell}
    \end{align}
    For randomized deign \ref{jna-choice-3:stratified-w-replace}, we have
    \begin{align}
        \ep_J[&\Gamma_0(u)]
        :=
        \frac{2u^2+1}{6\xi_1^3}\ep[g_1^3(X_1)]
        \notag\\
        &+
        \frac{(r-1)\big\{ (r^3+2r^2-2)u^2 + r^3-2r^2+2 \big\}}{2r^3\xi_1^3}\ep[g_1(X_1)g_1(X_2)g_2(X_1,X_2)]
        \label{def::jna-J1-Gamma_0}
        \\
        \ep_J[&\Gamma_\ell(u)]
        \textrm{ is the same as the $\ep_J[\Gamma_\ell(u)]$ under \ref{jna-choice-1:SRS-w-replace}}.
        \label{def::jna-J1-Gamma_ell}
    \end{align}
    In both examples, set
    \begin{align}
        G_J(u)
        :=&~
        \Phi(u)
        +
        \phi(u)
        \Big\{
            \frac{\ep_J[\Gamma_0(u)]}{\sqrt n}
            +
            \sum_{\ell=1}^{\lfloor\frac{\alpha/2}{\alpha-1}\rfloor}
            \frac{\ep_J[\Gamma_\ell(u)]}{\ma^\ell}
        \Big\},
    \end{align}
    where
    \begin{align}
        \ma
        :=&~
        n^{\alpha-1}\cdot \big\{1+1/(rn^{\alpha-1})\big\}
        \Big/
        \begin{cases}
            1+\dfrac{n^{\alpha-2}\xi_2^2\cdot r(r-1)}{\sum_{k=2}^r\binom{r}k \xi_k^2},
            & 
            \textrm{under \ref{jna-choice-1:SRS-w-replace}}
            \\
            1+\dfrac{n^{\alpha-2}\xi_2^2\cdot r^2(r-1)}{2\sum_{k=2}^r\binom{r}k \xi_k^2},
            & 
            \textrm{under \ref{jna-choice-3:stratified-w-replace}}
        \end{cases}.
    \end{align}
    Then we have
    \begin{align}
        \big\|
            F_{T_J+\delta_J}(u) - G_J(u)
        \big\|_\infty
        =&~
        O(n^{-\alpha/2}\log n).
        \label{eqn::corollary::application-jna-J1-J3::population-Edgeworth}
    \end{align}
\end{corollary}

We can naturally define the empirical version $\tilde G_J(u)$ with coefficient estimated by \eqref{def::xi_1-hat}, \eqref{def::g_1^3-hat}, \eqref{def::g_1g_1g_2-hat} and \eqref{est::our-design-sigma-h} and use it for downstream analysis, accompanied by theoretical guarantees exactly similar to Corollary \ref{corollary::one-sample::inference}.
We skip the repetitive details.

\section{Reduction of degenerate noiseless U-statistics}
\label{sec::one-sample::degenerate}

We call a U-statistic \emph{degenerate up to degree-$(k_0-1)$} \citep{chen2019randomized}, if for some $k_0\in[2:r]$, we have $\xi_1=\cdots=\xi_{k_0-1}=0$ 
and $\xi_{k_0}^2>0$. 
Unlike the non-degenerate scenario,
the incomplete U-statistics with deterministic and randomized designs are mainly driven by very different sources of stochastic variations, so they need to be treated separately.

We first analyze $U_J$ with the deterministic design $\jna$ in Section \ref{subsec::our-reduction-design}, starting with variance estimation.
Since this $\jna$ has the convenient property that $\sum_{I_k\in{\cal C}_n^k}\ank{k}^2(I_k) = \binom rk |\jna|$ for all $k\in[1:r]$, we have $\var(U_J) = |\jna|^{-1}\sum_{k=k_0}^r \binom rk \xi_k^2 = |\jna|^{-1} \sigma_h^2$, where $\sigma_h^2:=\var(h(X_{[1:r]})) = \sum_{k=k_0}^r\binom{r}k \xi_k^2$ can be accurately estimated in $O(n^\alpha)$ time by
$\tilde\sigma_h^2:=
|\jna|^{-1}\sum_{I_r\in\jna}
\Big\{
    h\big(X_{I_r}\big) - U_J
\Big\}^2$.
We studentize $U_J$ as $T_J:=(U_J-\mu)/\big(|\jna|^{-1/2}\cdot \tilde \sigma_h\big)$.
\begin{theorem}
    \label{main-theorem::one-sample::degenerate::deterministic::normality}
    Suppose $h$ is degenerate up to degree $(k_0-1)$ for some $k_0\in[2:r]$.  
    Set $\alpha\in (1,2)$.
    If $\jna$ is constructed as that in Section \ref{subsec::our-reduction-design}, then as $n\to\infty$,
    $
        T_J
        \stackrel{d}\to
        N(0,1)
    $.
\end{theorem}
The downstream inference using Theorem \ref{main-theorem::one-sample::degenerate::deterministic::normality} is completely standard.  We skip it.

Next, we analyze $U_J$ with a randomized $\jna$.  
For simplicity, we focus on two particular designs \ref{jna-choice-1:SRS-w-replace} and \ref{jna-choice-3:stratified-w-replace}, where $\jna$ elements are independent.
The decomposition \eqref{analysis::randomized-incomplete-U-stat-conventional} from \citet{chen2019randomized} now becomes particularly useful.
In the degenerate scenario, $V_J$ is the dominating term in \eqref{analysis::randomized-incomplete-U-stat-conventional} whenever $\alpha<k_0$.
By \citet{major2007}, we know $U_n-\mu=\tOp\big( n^{-k_0/2}\log^{k_0/2}n \big)$.  
Rewrite $V_J$ as
$V_J = |\jna|^{-1} \sum_{j=1}^{|\jna|}{\cal X}_j$,
where ${\cal X}_j := h(X_{I_r^{(j)}})-U_n$ by recalling \eqref{def::jna}.
Conditioning on $X_{[1:n]}$, we can view $V_J$ as a sample mean of independent mean-zero random variables.
We have
$
|\jna|\cdot \var(V_J|X_{[1:n]})
    =
    \binom nr^{-1} \sum_{I_r\in{\cal C}_n^r} \big\{ h(X_{I_r}) - U_n \big\}^2
    =
    \sigma_h^2 + \tOp\big( n^{-k_0/2}\log^{k_0/2}n \big)
$. 
Estimating $\sigma_h^2$ by $\tilde\sigma_h^2$,
the studentization is
\begin{align}
    T_J
    :=
    (U_J-\mu)/\big(|\jna|^{-1/2}\cdot \tilde \sigma_h\big)
    =
    \tOp\big( n^{(\alpha-k_0)/2}\log^{k_0/2}n \big)
    +
    V_J/(|\jna|^{-1/2}\cdot \tilde \sigma_h).
    \label{degenerate::random-design::studentization}
\end{align}
To characterize the distribution of $T_J$, we study the second term on the RHS of \eqref{degenerate::random-design::studentization}.
We formulate an Edgeworth expansion that refines the normal approximation \citep{chen2019randomized}.
Define
\begin{align}
    G_{V|X_{[1:n]}}(u)
    :=&~
    \Phi(u)
    +
    \frac{\ep[{\cal X}_1^3|X_{[1:n]}] (2u^2+1)\phi(u)}
    {\big\{\var({\cal X}_1|X_{[1:n]})\}^{3/2} \cdot 6 |\jna|^{1/2}}
    \\
    \tilde G_{V|X_{[1:n]}}(u)
    :=&~
    \Phi(u)
    +
    \frac{\tilde\nu_h^3 (2u^2+1)\phi(u)}
    {\tilde\sigma_h^3 \cdot 6 |\jna|^{1/2}},
\end{align}
and define
$
    \tilde\nu_h^3
    :=
    |\jna|^{-1} \sum_{I_r\in\jna}\big\{ h(X_{I_r}) - U_J \big\}^3
$ as the empirical version of $\ep[{\cal X}_1^3|X_{[1:n]}]$. 
The next theorem bounds the distribution approximation errors of $G_{V|X_{[1:n]}}(u)$ and $\tilde G_{V|X_{[1:n]}}(u)$.
\begin{theorem}
    \label{main-theorem::one-sample::degenerate::random::normality}
    Suppose the U-statistic is degenerate up to degree $(k_0-1)$ for some $k_0\in[2:r]$.  
    Set $\alpha\in (1,k_0)$.
    If $\jna$ is follows \ref{jna-choice-1:SRS-w-replace} or \ref{jna-choice-3:stratified-w-replace} in Section \ref{subsec::analyasis-incomplete-u-stat}, then
    \begin{align}
        \Big\|
            F_{T_J+\delta_V|X_{[1:n]}}(u)
            -
            G_{V|X_{[1:n]}}(u)
        \Big\|_\infty
        =&~
        \tOp\Big(
            n^{-\alpha}\log^{1/2}n
            +
            n^{(\alpha-k_0)/2}\log^{k_0/2}n
        \Big)
        \label{eqn::degenerate::Edgeworth-population}
        \\
        \Big\|
            F_{T_J+\delta_V|X_{[1:n]}}(u)
            -
            \tilde G_{V|X_{[1:n]}}(u)
        \Big\|_\infty
        =&~
        \tOp\Big(
            n^{-\alpha}\log^{1/2}n
            +
            n^{(\alpha-k_0)/2}\log^{k_0/2}n
        \Big)
        ,
        \label{eqn::degenerate::Edgeworth-empirical}
    \end{align}
    where we add to $T_J$ an artificial Gaussian smoother independent of all data $\delta_V\sim N(0, \sigma^2=C n^{-\alpha}\log n)$ with a large enough constant $C$.
\end{theorem}
To construct the Cornish-Fisher CI, we invert $\tilde G_{V|X_{[1:n]}}(u)$.
\begin{corollary}
    \label{corollary::one-sample::degenerate::random::CI}
    Define
    \begin{align}
        \tilde G_{V|X_{[1:n]}}^{-1}(z_\beta)
        :=&~
        z_\beta
        -
        \frac{\tilde\nu_h^3 (2z_\beta^2+1)}
        {\tilde\sigma_h^3 \cdot 6 |\jna|^{1/2}}.
    \end{align}
    and the two-sided Cornish-Fisher confidence interval for estimating $\mu$ to be
    \begin{align}
        {\cal I}_V
        :=
        \Big(&
            U_J
            -
            \big\{
                \tilde G_{V|X_{[1:n]}}^{-1}(1-z_\beta/2)
                -
                \delta_V
            \big\}
            |\jna|^{-1/2} \cdot \tilde \sigma_h
            ,
            \notag\\
            &
            U_J
            -
            \big\{
                \tilde G_{V|X_{[1:n]}}^{-1}(z_\beta/2)
                -
                \delta_V
            \big\}
            |\jna|^{-1/2} \cdot \tilde \sigma_h
        \Big),
    \end{align}
    Then we have
    \begin{align}
        \pr(\mu\in {\cal I}_V)
        =&~
        1-\beta 
        + 
        O\big(
            n^{-\alpha}\log^{1/2}n + n^{(\alpha-k_0)/2}\log^{k_0/2}n
        \big).
    \end{align}
\end{corollary}

Theorem \ref{main-theorem::one-sample::degenerate::random::normality} and Corollary \ref{corollary::one-sample::degenerate::random::CI} provide the first finite-sample characterization of the trade-off between risk control accuracy and computational speed in the reduction of degenerate U-statistics.
We remark three quick interpretations of these results.
First, when $\alpha<k_0/2$, our method achieves a sharper error bound than normal approximation.
Second, our method's error rate is optimized to $O\big( n^{-k_0/3}\log^{k_0/2}n \big)$ at $\alpha=k_0/3$.
Third, for all choices of $\alpha<r$, our method's error bound improves over the  $O\big(n^{(\alpha-k_0)/4}+n^{-(\alpha\xiao 1)/6}\big)$ Berry-Esseen bound  \citep{chen2019randomized} for one-dimensional U-statistics.\footnote{The first term in the error bound of Theorem 3.3 in \citet{chen2019randomized} needs a slight revision, changing ``$\log^{k+3} d$'' into ``$\log^{k+3} (d+C_0)$'' for some constant $C_0>0$.  To see this, notice that when (the ``$d$'' in their paper) $d=1$, Equation (C.19) in the proof of Theorem 3.3 \citep{chen2019randomized-SUPP} needs this modification to hold.}

We conclude this subsection by an interesting comparison between the guidance on selecting $\alpha$ in the non-degenerate and degenerate scenarios.
For non-degenerate U-statistics, the reason not to choose a large $\alpha$ is purely computation (recall Remark \ref{remark::noiseless-why-not-alpha-greater-than-2}); 
if computation resources permit, we can set $\alpha=r$ to minimize $\var(U_J)$.
The degenerate scenario, however, is very different.
As $\alpha$ approaches $k_0$, the risk control accuracy of both our method and normal approximation will drop, because the \emph{non-Gaussian} part $U_n-\mu$ becomes increasingly non-ignorable in \eqref{analysis::randomized-incomplete-U-stat-conventional}. 
This blessing reinstatement of normality by incompleteness was discovered by \citet{weber1981incomplete, chen2019randomized}, and our paper provides a much-refined finite-sample description of this phenomenon.
Approximating the distribution of the non-Gaussian $U_n-\mu$ is quite challenging, and no finite-sample result yet exists to our best knowledge \citep{van2000asymptotic,giraudo2021limit}.

\section{Reduction of network moments as noisy U-statistics}
\label{section::network-U-stat}

Sections \ref{section::our-method} and \ref{sec::one-sample::degenerate} comprehensively addressed noiseless U-statistics.
Now, we turn our attention to an important class of noisy U-statistics, called \emph{network moments}, for which the data-oblivious reduction schemes is particularly useful.  
Consider an undirected network with binary edges, where each node is associated with a latent position $X_i\sim$~Uniform$[0,1]$, and the edge probability between $(i,j)$ is $W_{i,j} := \rho_n\cdot f(X_i,X_j)$ \citep{aldous1981representations, hoover1979relations}, where $\rho_n$ is a global sparsity parameter and $f(\cdot,\cdot)$ is a latent function, called \emph{graphon function}, that encodes all network structural information.  Then observed data are $A_{i,j}|W\stackrel{\rm i.i.d.}\sim$~Bernoulli$(W_{i,j})$.  
A network moment, indexed by a motif $R$ of $r$ nodes and $s$ edges, is the count of $R$ in the network.  Specifically, let $h(A_{[1:r]}):=\sum_{\pi\in{\rm Permu.([1:r])}}\mathbbm{1}_{A_{[\pi(1:r)]}\geq R}$, in which we treat $R$ as its own adjacency matrix, and $\geq$ is element-wise. 
Define the empirical network moment $\hat U_n$ by
\begin{align}
    \hat U_n
    :=&~
    \binom{n}r^{-1} \sum_{I_r\in {\cal C}_n^r} h(A_{I_r}).
    \label{def::network-U-stat}
\end{align}
Denote $\mu:=\ep[U_n]$.  The goal is to perform statistical inference for $\mu$.

The higher-order approximation to the distribution of the studentization of the complete network moment $\hat U_n$ has been established in \citet{zhang2020edgeworth}.
However, similar to the noiseless scenario, the expensive $O(n^r)$ computation complexity is also a major obstacle in statistical inference.
This leads us to consider U-statistic reduction.
Notice that in \eqref{def::network-U-stat}, both $f$ and $X_i$'s are not only unobserved, but even inestimable, due to an identifiability issue \citet{gao2015rate}.
Therefore, data-aware reduction techniques such as \citep{kong2020design} are inapplicable, leaving data-oblivious reduction the natural choice.
The incomplete network U-statistic with design $\jna$ is
\begin{align}
    \hat U_J
    :=
    \big|\jna\big|^{-1}
    \sum_{I_r\in \jna}h(A_{I_r})
    =
    \big|\jna\big|^{-1}
    \sum_{I_r\in {\cal C}_n^r} \ank{r}(I_r) h(A_{I_r}).
    \label{def::hat-U_J}
\end{align}
Like \citet{zhang2020edgeworth}, we decompose the stochastic variations in $\hat U_J$ into two parts:
$
    \hat U_J - \mu = (U_J-\mu) + (\hat U_J - U_J)
$,
where we define
$
    U_J := \ep[\hat U_J|X_{[1:n]}]
    = 
    \mu 
    + 
    |\jna|^{-1}\sum_{i=1}^n \ank1(i)g_1(X_i) 
    + 
    |\jna|^{-1}\sum_{k=2}^r\sum_{I_k\in{\cal C}_n^k} \ank{k}(I_k)g_k(X_{I_k})
$.
We call a network moment $\hat U_J$ non-degenerate if the noiseless part in its decomposition $U_J$ is non-degenerate.
In this paper, we focus on non-degenerate network moments, leaving the degenerate case to an interesting future work.

To design a variance estimator for studentizing $\hat U_J$, notice that $\var(U_J-\mu|\jna)\asymp \rho_n^{2s}n^{-1}$ while $\var(\hat U_J-U_J|\jna)\asymp \rho_n^{2s-1}n^{-2}+\rho_n^s n^{-\alpha}$ (see proof of Theorem \ref{theorem::main-theorem::network-U-stat}).
When $\rho_n^s n^{\alpha-1}\to \infty$, the randomness in $U_J$ dominates that in $A|W$\footnote{Notice that this condition is always satisfied for \emph{complete} network U-statistics, see \citet{zhang2020edgeworth}:  where $\alpha=r$ (i.e. $U_J=U_n$), with the presence of Condition (ii) in their Lemma 3.1.}.
We now assume $\rho_n^s n^{\alpha-1}\to\infty$ and relegate the case $\rho_n^s n^{\alpha-1}\to 0$ to Theorem \ref{theorem::main-theorem::network-normality}\footnote{For simplicity, we only study the case where $\rho_n^s n^{\alpha-1}\to 0$, leaving the intermediate case to future work.}.
With this assumption, we can approach $\var(\hat U_J)$ via $\xi_1^2$.  
Define $\hat a_i:=\sum_{I_r\in\jna, i\in I_r}h(A_{I_r})/\ank1(i)$, and estimate $\xi_1^2$ by
\begin{align}
    \hat\xi_1^2
    :=&~
    r^{-1}|\jna|^{-1}\sum_{i=1}^n \ank1(i)
    \big(
        \hat a_i - \hat U_J
    \big)^2.
    \label{def::network-xi-hat}
\end{align}
There are several good alternative choices for $\hat\xi_1^2$, e.g. $\big\{\sum_{i=1}^n \ank1^2(i)\big\}^{-1}\sum_{i=1}^n \ank1^2(i) (\hat a_i-\hat U_J)^2$ and $n^{-1}\sum_{i=1}^n (\hat a_i-\hat U_J)^2$; we choose \eqref{def::network-xi-hat} for technical convenience.  
Studentize $\hat U_J$ as
\begin{align}
    \hat T_J
    :=
    \frac{\hat U_J - \mu}{\{\sum_{i=1}^n \ank1^2(i)\}^{1/2} \cdot \hat\xi_1 / |\jna|}
    +
    \delta_J,
\end{align}
where, like before, we add an artificial noise $\delta_J$ to completely waive Cram\'er's condition.
For a motif $R$, define the population Edgeworth expansion for $\hat T_J$ as follows.
\begin{align}
    G_J(x)
    :=&~
    \Phi(x) + \frac{\varphi(x)}{\sqrt n\cdot \xi_1^3}
    \Big\{
        \frac{2x^2+1}{6} \cdot\ep[g_1^3(X_1)]
        \notag\\
        &+
        \frac{r-1}{2}\cdot (x^2+1) \ep[g_1(X_1)g_1(X_2)g_2(X_1,X_2)]
    \Big\}.
    \label{def::network-population-edgeworth}
\end{align}
To formulate the empirical Edgeworth expansion (EEE), we also need to estimate $\ep[g_1^3(X_1)]$ and $\ep[g_1(X_1)g_1(X_2)g_2(X_1,X_2)]$.  Limited by the $O(n^\alpha)$ computational budget, we need to estimate these quantities very differently than that in \citet{zhang2020edgeworth}.  Define
\begin{align}
    \hat\ep[g_1^3(X_1)]
    :=&~
    r^{-1}|\jna|^{-1}\sum_{i=1}^n \ank1(i) \big(\hat a_i-\hat U_J\big)^3,
    \\
    \hat\ep[g_1(X_1)g_1(X_2)&g_2(X_1,X_2)]
    \notag\\
    :=&~
    |\jna|^{-1}
    \sum_{I_r\in\jna} 
    \big\{(h(A_{I_r})-\hat U_J)
    \cdot \big(\hat a_{I_r(1)} - \hat U_J\big)
    \cdot \big(\hat a_{I_r(2)} - \hat U_J\big)\big\},
\end{align}

where $I_r(\ell)$ indicates the $\ell$th element of $I_r$, and define the EEE, denoted by $\hat G_J(u)$, to be the empirical version of $G_J(u)$ with $\xi_1$, $\ep[g_1^3(X_1)]$ and $\ep[g_1(X_1)g_1(X_2)g_2(X_1,X_2)]$ replaced by their estimators.  Define
\begin{align}
    \tilde M_\alpha
    (\rho_n,n,\alpha;R)
    :=&~
    \begin{cases}
        \rho_n^{-1}n^{-(\alpha-1)},
        &
        (r=2)
        \\
        \Big\{
            \rho_n^{-s/2}n^{-(\alpha-1)/2}
            +
            \max_{p\in [3:(\lceil\alpha\rceil-1)]} \rho_n^{-v_p/2} n^{-(p-1)/2}
        \Big\}
        \log^{1/2}n,
        &
        (r\geq 3)
    \end{cases}
    \notag\\
    & +
    n^{-1}\cdot \log^{3/2}n
    +
    \begin{cases}
            (\rho_n\cdot n)^{-1}
        \cdot \log^{1/2}n,
        & \textrm{ (Acyclic }R)\\
            \rho_n^{-r/2}\cdot n^{-1}
        \cdot \log^{1/2}n,
        & \textrm{ (Cyclic }R)
    \end{cases}
    ,
\end{align}
where $v_p$ is the maximum number of edges among all $p$-node subgraphs of $R$.
We have
\begin{theorem}
    \label{theorem::main-theorem::network-U-stat}
    Suppose the network is generated by the graphon model.
    Set $\alpha\in (1,r]$ satisfying 
    $\tilde M_\alpha(\rho_n,n,\alpha;R)\to 0$.
    Also assume the following conditions:
    \begin{enumerate}
        \item $\rho_n^{-2s}\cdot \xi_1^2\geq {\rm constant}>0$.
        \item The design $\jna$ satisfies Assumptions \ref{New-assumption-1} and \ref{New-assumption-2}.
    \end{enumerate}
    Then we have
    \begin{align}
        \big\|
            F_{\hat T_J(u)} - G_J(u)
        \big\|_\infty
        =&~
        O(\NEWnetworkerror),
        \\
        \big\|
            F_{\hat T_J(u)} - \hat G_J(u)
        \big\|_\infty
        =&~
        \tOp(\NEWnetworkerror).
    \end{align}
    
\end{theorem}

Theorem \ref{theorem::main-theorem::network-U-stat} provides the first rigorous description of the trade-off between computational speed and the accuracy of risk control in reduced network moments, reflected by the first term in $\NEWnetworkerror$.  
Theorem \ref{theorem::main-theorem::network-U-stat} sharply contrasts Theorem \ref{theorem::general-main-theorem}.
For noiseless U-statistics, we can achieve higher-order accurate risk control by setting $\alpha>1$, and increasing $\alpha$ beyond 2 will not further benefit this accuracy; but for network moments with $r\geq 3$ edges (i.e. except $R=$~Edge), even in the densest network setting $\rho_n\asymp 1$, higher-order accurate risk control still requires $\alpha>2$ regardless of the motif's shape.
This distinction is caused by edge-wise observational errors.
A close inspection of the result of Theorem \ref{theorem::main-theorem::network-U-stat} also finds an interesting fact that for completely dense networks, increasing computational budget beyond $O(n^3)$ will not further improve inference accuracy, analogous to its sibling result for noiseless U-statistics.
Comparing the results of Theorem \ref{theorem::main-theorem::network-U-stat} and Theorem \ref{theorem::general-main-theorem} also raises another interesting question: if we incorporate some $O(n^{-(\alpha-1)\ell})$ terms, like those in \eqref{Edgeworth::generic}, into \eqref{def::network-population-edgeworth}, will that improve the error bound in Theorem \ref{theorem::main-theorem::network-U-stat}?
The answer is no, because the first term in $\NEWnetworkerror$ is \emph{not} caused by ignoring those terms in the Edgeworth expansion, but instead due to edge-wise observational errors.

\begin{figure}[h!]
    \centering
    \makebox[\textwidth][c]{
    \includegraphics[width=0.25\textwidth]{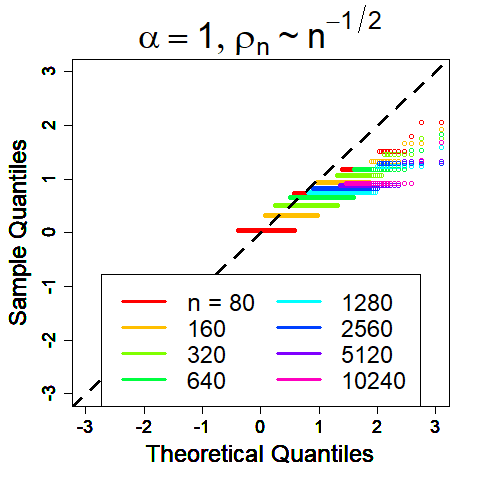} 
    \includegraphics[width=0.25\textwidth]{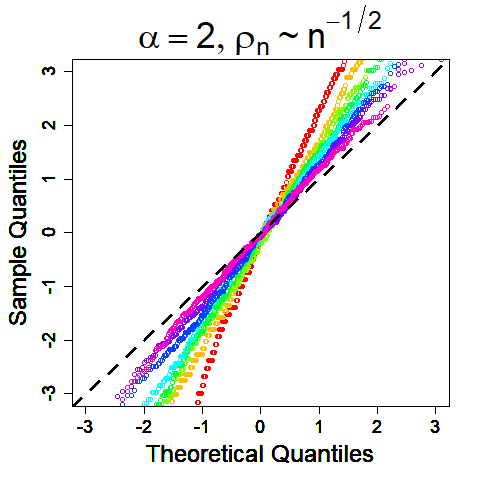} 
    \includegraphics[width=0.25\textwidth]{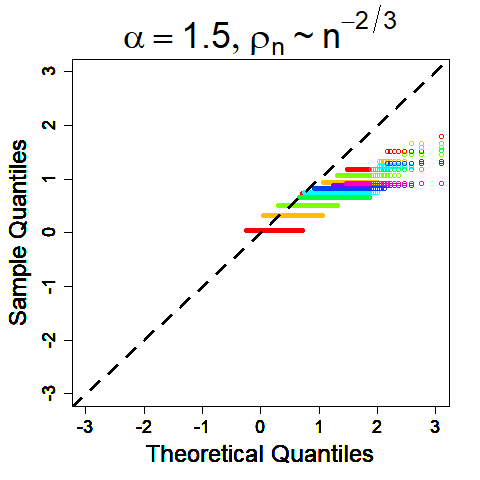} 
    \includegraphics[width=0.25\textwidth]{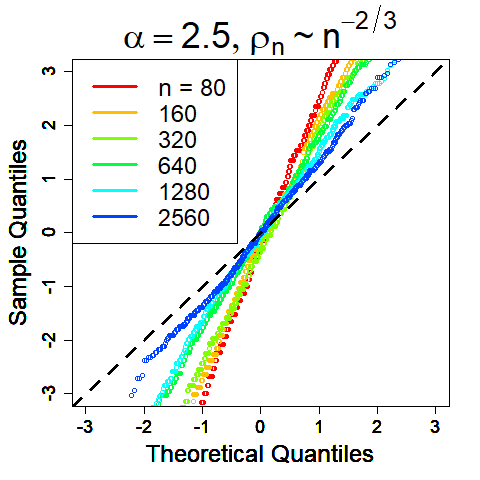} 
    }
    \caption{
    Network moments (sparse networks):
    QQ plot of sampling distribution of $\hat T_J'$,
    Triangle, {\tt BlockModel}.
    Columns 1 \& 2: $\rho_n \asymp n^{-1/2}$; 
    columns 3 \& 4: $\rho_n \asymp n^{-2/3}$. 
    Due to memory limit, we did not run $n\geq 5120$ for $\alpha=2.5$.
    }
    \label{simulation::sparser-network-qqnorm-2}
\end{figure}

We conclude this section by studying the case $\rho_n^s n^{\alpha-1}\to0$, either due to network sparsity, or because we reduce the computation too much.
Now, the randomness in $\hat U_J$ is dominated by the randomness in $\hat U_J - U_J$.
Not surprisingly, the variance estimator for $\hat U_J$ needs some revision.
With the new variance estimator
$
    (\hat\sigma_J')^2
    :=
    |\jna|^{-1}\hat U_J
$,
the studentization form is
\begin{align}
    \hat T_J'
    :=&~
    \frac{\hat U_J-\mu}{\hat\sigma_J'}.
    \label{def::T_J'::very-sparse-networks}
\end{align}
The next theorem establishes the asymptotic normality of $\hat T_J'$.
Define $\check M_\alpha(\rho_n,n,\alpha;R)$ to almost equal $\NEWnetworkerror$, except for replacing $\rho_n^{-s/2}n^{-(\alpha-1)/2}$ by $\rho_n^{-s/2}n^{-\alpha/2}$.
\begin{theorem}
    \label{theorem::main-theorem::network-normality}
    Assume the conditions of Theorem \ref{theorem::main-theorem::network-U-stat} hold, except replacing the condition $\NEWnetworkerror\to 0$
    by $\check M_\alpha(\rho_n,n,\alpha;R)\to 0$. 
    Then we have
    \begin{align}
        \hat T_J' \stackrel{d}\to N(0,1).
    \end{align}
\end{theorem}
In Theorem \ref{theorem::main-theorem::network-normality}, we still need the assumption $\rho_n^s n^\alpha\to\infty$, which is not much weaker than $\rho_n^s n^{\alpha-1}\to\infty$, but based on the empirical evidence from our simulation result (Section \ref{simulation:network-U-stat}, Figure \ref{simulation::sparser-network-qqnorm-2}), we conjecture that the assumption $\rho_n^s n^\alpha\to\infty$ should be minimal\footnote{As a theoretical evidence: consider the simplest motif: $R=$~Edge.  When $\rho_n\cdot n^\alpha\to \lambda$ (constant), the limiting distribution of $\hat T_J'$, after rescaling, is Poisson, not normal \citep{novak2019poisson}.}.

\section{Simulations}
\label{section::simulations}

\subsection{Simulation 1: non-degenerate U-statistics}
\label{simulation::simulation-1}
Our first simulation assesses CDF approximation accuracy for noiseless non-degenerate U-statistics.
The goal is to accurately approximate $F_{ T_J+\delta_J}$, where we set a small variance with $C_\delta = 0.008$ for $\delta_J$.
We generated synthetic data with $X_1,\ldots,X_n\stackrel{\rm i.i.d.}\sim $~PDF: $(x+1)/2$, $x\in [-1,1]$; and the kernel function $h(x_1,x_2,x_3) := \sin(x_1+x_2+x_3)$.
We experimented our proposed deterministic design in Section \ref{subsec::our-reduction-design} and a random design \ref{jna-choice-1:SRS-w-replace} from Section \ref{subsec::analyasis-incomplete-u-stat}.
We compared our method to the following benchmarks:
1. $N(0,1)$; 2. Resample bootstrap (bootstrap iteration $B=200$ \citep{levin2019bootstrapping}); 3. Subsample bootstrap (subsample size: $n^{1/2}$).
We emulate the true sampling distribution of $T_J + \delta_J$ by a Monte-Carlo approximation with samples $n_{MC}:= 10^6$.  The performance measure is
\begin{align}
    \sup_{u\in[-2,2];10u\in\mathbb{Z}}
    \big|
        \hat{F}_{T_J+\delta_J}(u)- F_{T_J+\delta_J}(u)
    \big|.
    \label{simulation::non-degenerate::CDF::error-measure}
\end{align}
We varied $n \in \{10, 20, 40, 80\}$ and set $\alpha=1.5$ (result for $\alpha=1.7$ in Supplementary Material). 
For each $(n,\alpha)$ setting, we repeated the experiment 30 times and recorded the mean and standard deviation of the distribution approximation errors \eqref{simulation::non-degenerate::CDF::error-measure}.

\begin{figure}[h!]
    \centering
    \makebox[\textwidth][c]{
    \raisebox{0.3em}{
    \includegraphics[width=0.26\textwidth]{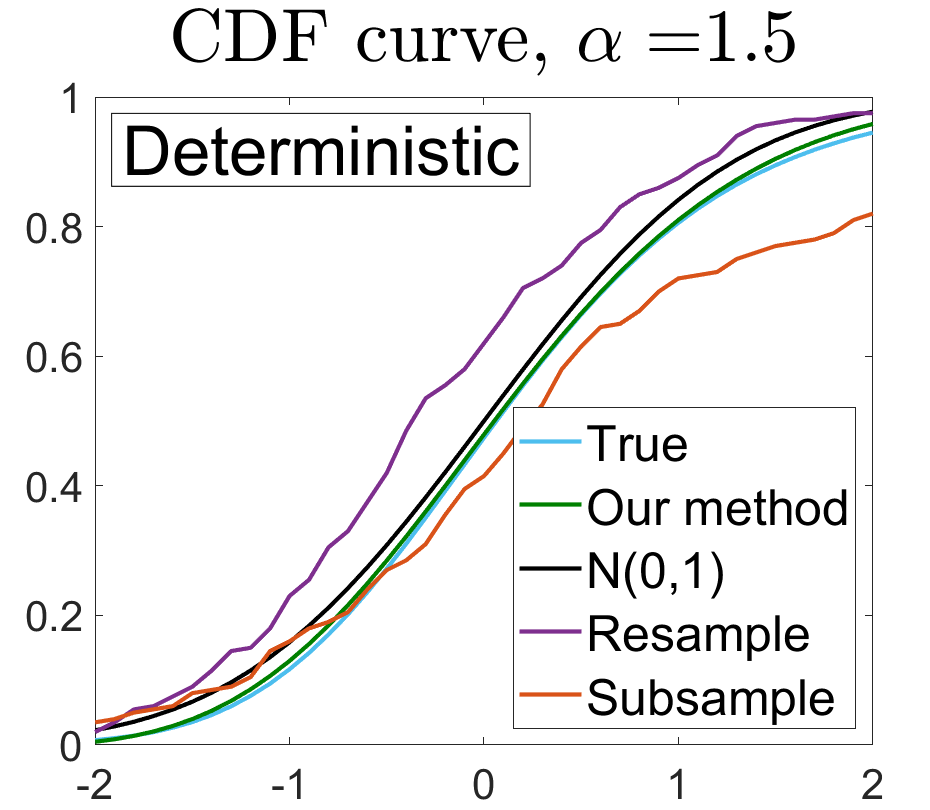}
    }
    \includegraphics[width=0.25\textwidth]{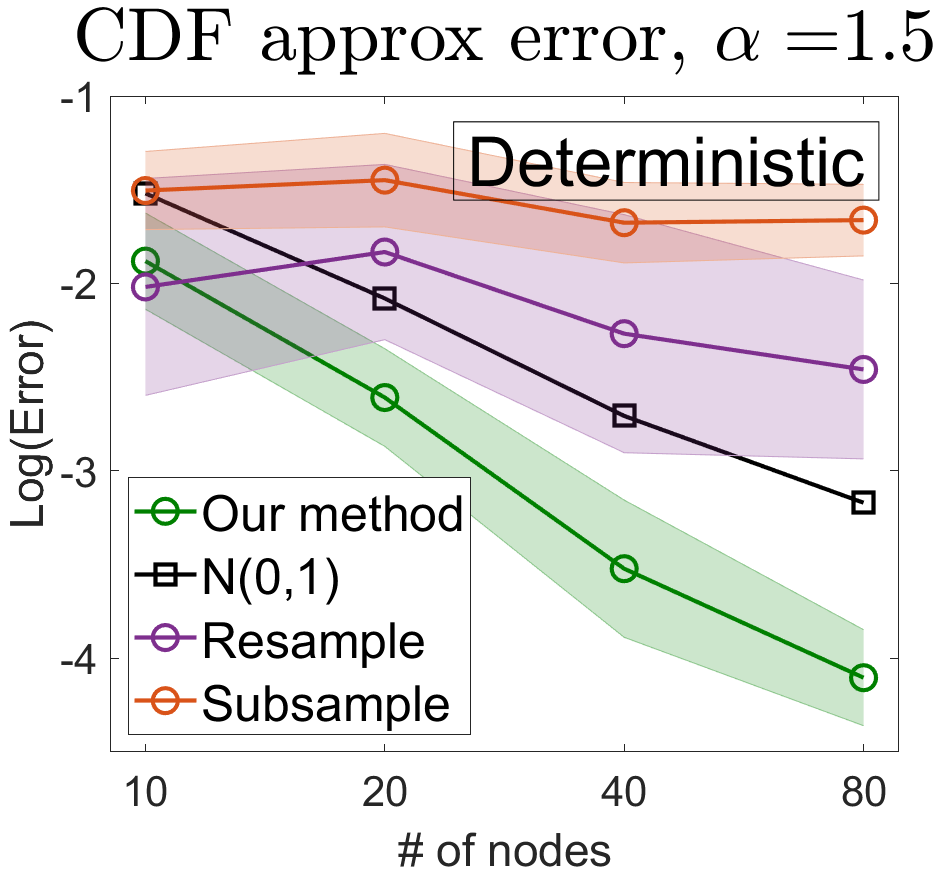}
    \includegraphics[width=0.25\textwidth]{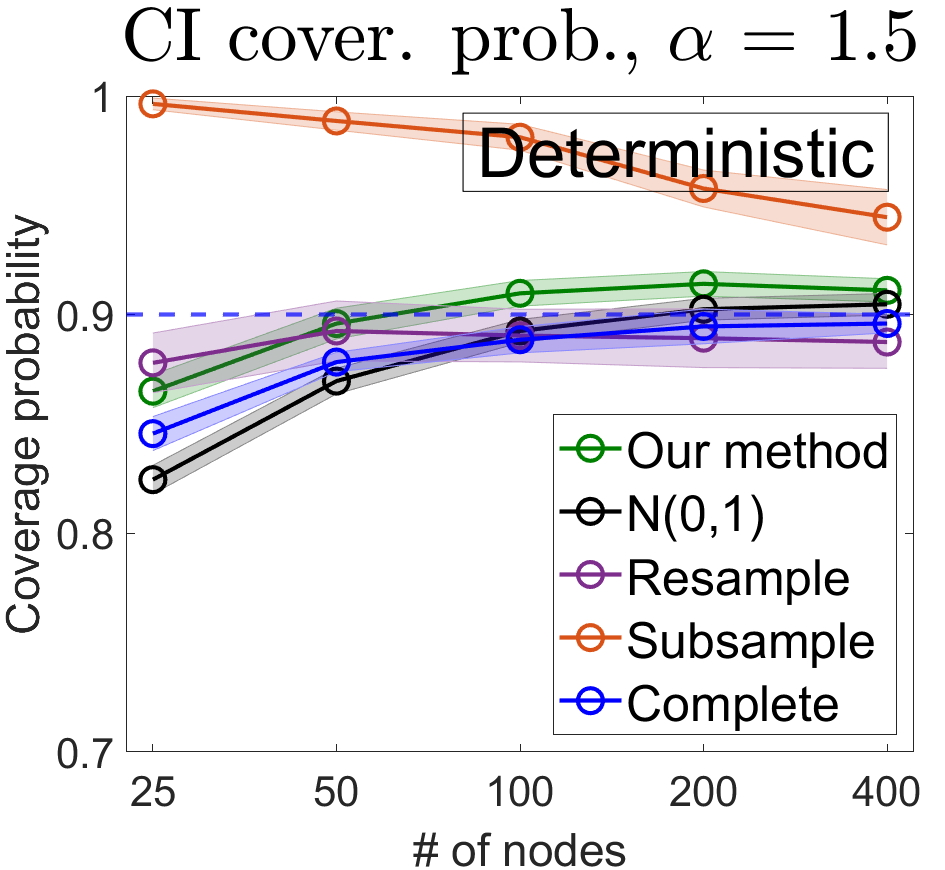}
    \includegraphics[width=0.25\textwidth]{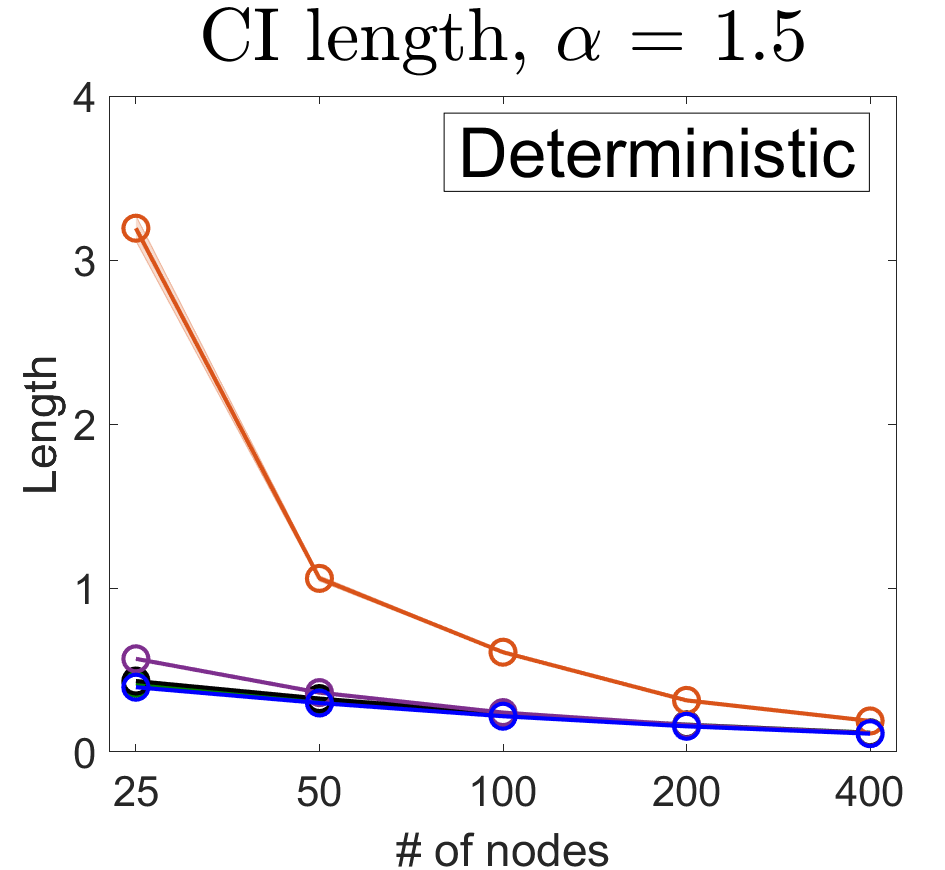}
    \includegraphics[width=0.25\textwidth]{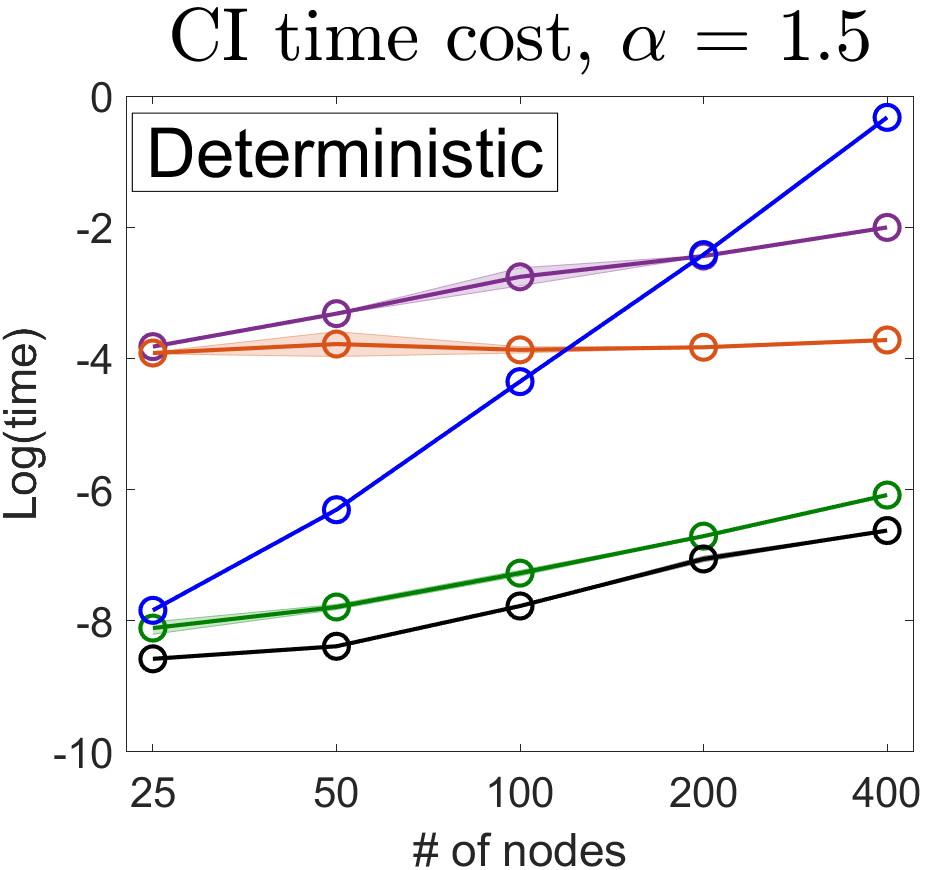}
    }
    \makebox[\textwidth][c]{

    \raisebox{0.3em}{
    \includegraphics[width=0.26\textwidth]{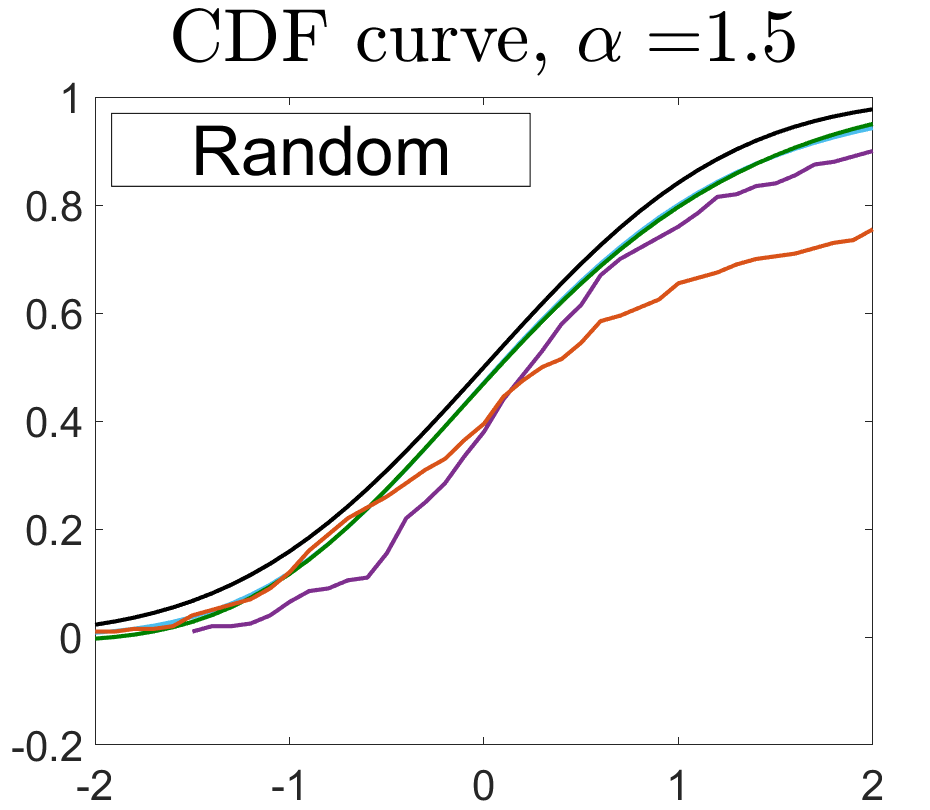}
    }
    \includegraphics[width=0.25\textwidth]{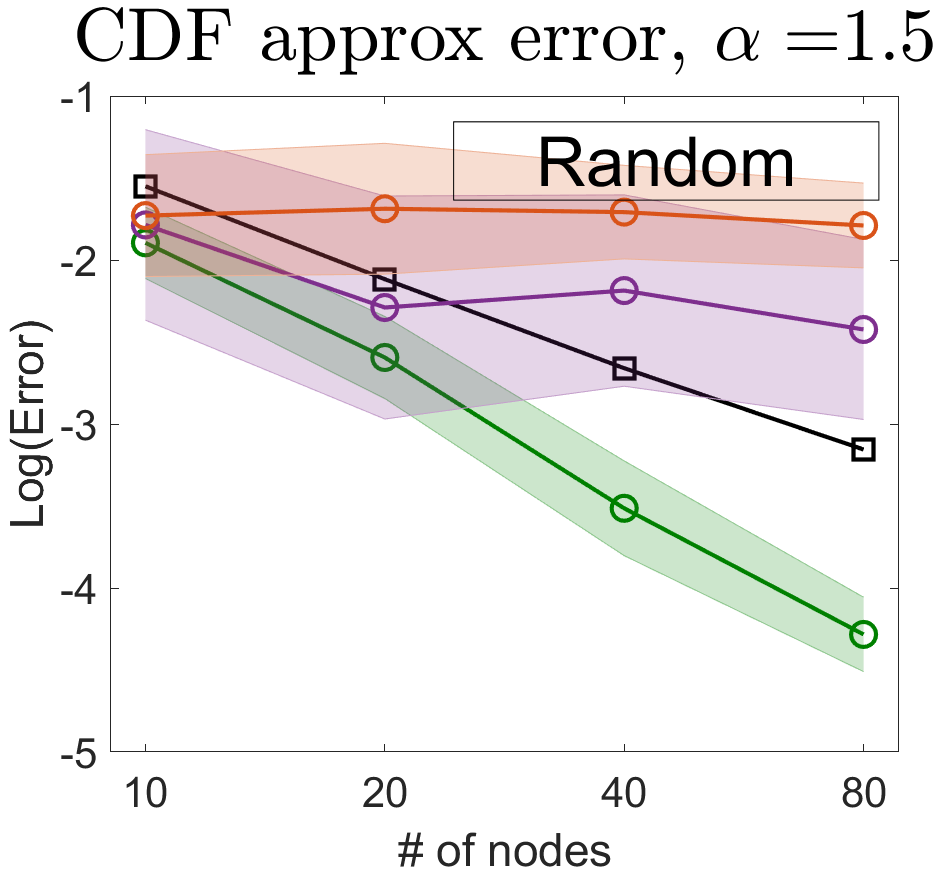}
    \includegraphics[width=0.25\textwidth]{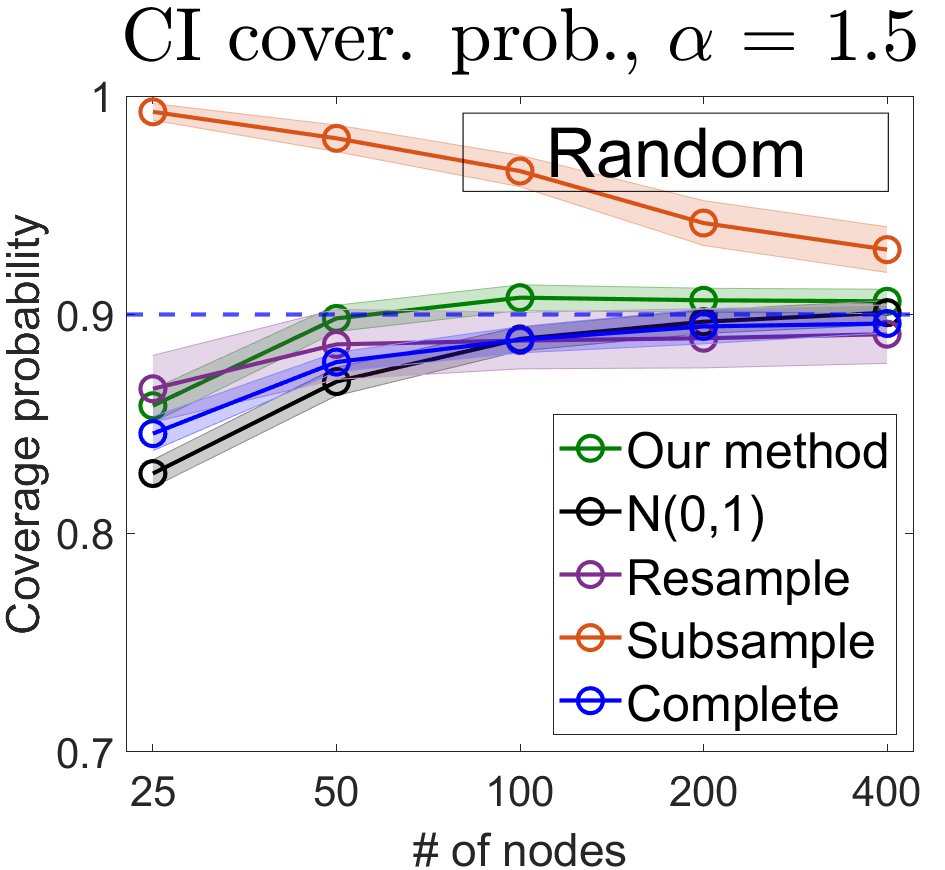}
    \includegraphics[width=0.25\textwidth]{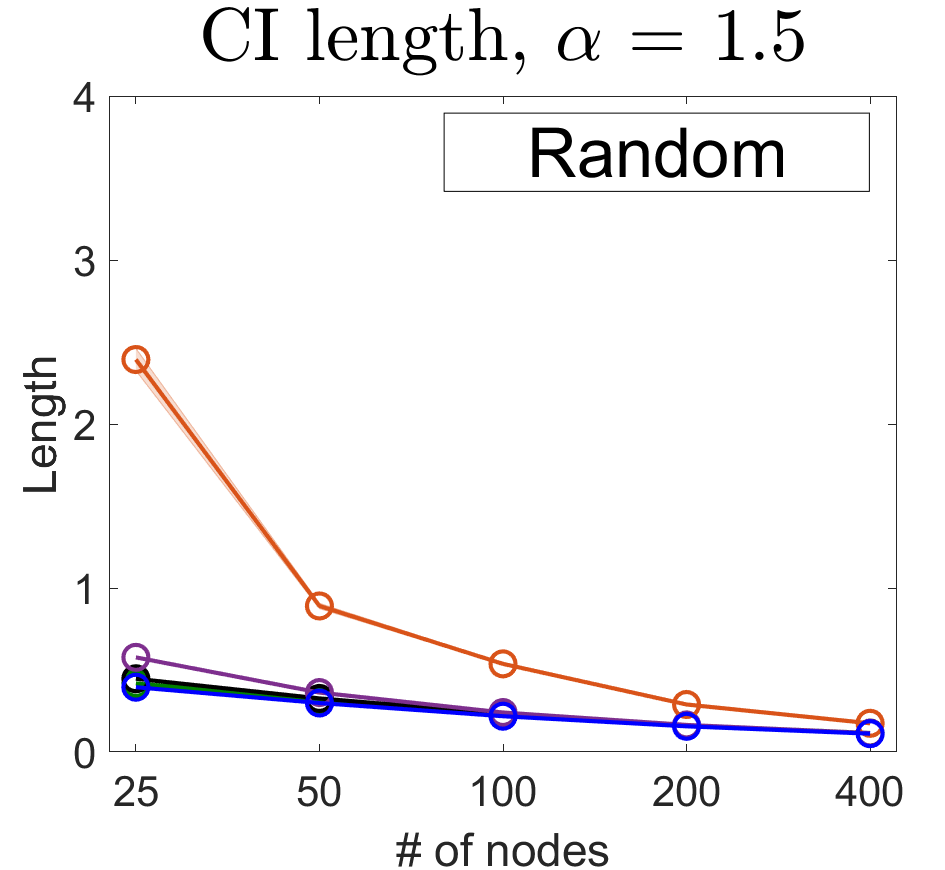}
    \includegraphics[width=0.25\textwidth]{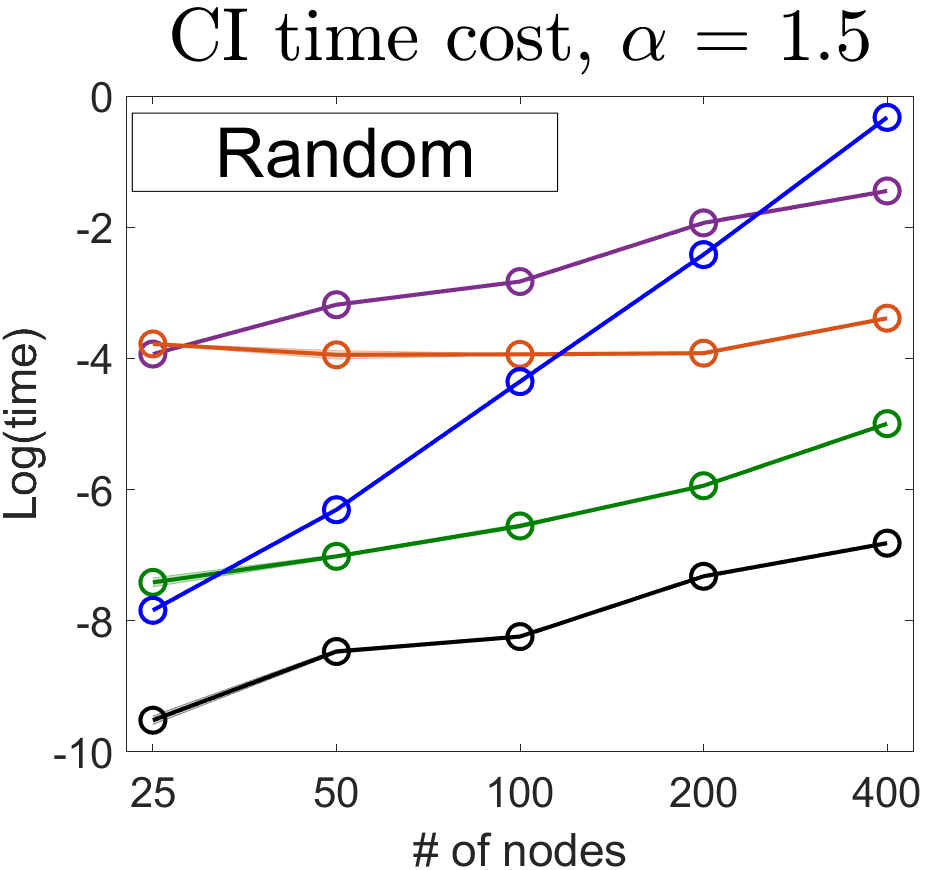}
    }
    
    \caption{
    Non-degenerate U-statistics:
    Row 1: deterministic $\jna$ (Section \ref{subsec::non-degenerate::applications}); 
    row 2: random $\jna$ (scheme \ref{jna-choice-1:SRS-w-replace}). 
    Column 1: true CDF~$=F_{T_J+\delta_J}(u)$, $n = 80$;
    column 2: CDF approximation error; 
    column 3: CI coverage probability,  dashed blue line~$=1-\beta=90\%$; 
    column 4: CI length; column 5: computation time.
    }
    \label{simulation::non-degenerate::cdf-approx}
\end{figure}

Column 1 of Figure \ref{simulation::non-degenerate::cdf-approx} shows the true and estimated CDF curves for $T_J+\delta_J$.  We can see that our method's estimated CDF curve almost overlaps the true CDF curve; whereas other methods all exhibit noticeable deviations at different levels.
Then Column 2 of Figure \ref{simulation::non-degenerate::cdf-approx} shows log-transformed CDF approximation errors for different methods under different $(n,\alpha)$ configurations.
Our method shows clear advantage in accuracy across all settings.
Specifically, our method is the only method with an empirical error rate faster than $n^{-1/2}$.
This well-matches our theoretical prediction on the higher-order accuracy of our method.

Next, we compare our Cornish-Fisher confidence interval to the three benchmark methods in Simulation 1, plus the C-F CI constructed based on the complete U-statistic, in the following aspects: 
coverage probability, CI length and computation time.
We fixed confidence level $1-\beta = 90\%$ and focused on two-sided confidence intervals for simplicity.
Most simulation settings were inherited from Simulation 1, but now we do not need a large $n_{MC}$ and can afford larger $n$'s, so we varied $n \in \{25, 50, 100, 200, 400\}$.
In each experiment, to evaluate one empirical CI coverage probability, we generate 3000 CI's for our method, $N(0,1)$ and complete U-statistics; and 500 CI's for resampling and subsampling bootstraps since they take longer to run; and the experiment is repeated 100 times for all methods except the complete U-statistic method (repeated 20 times).

Figure \ref{simulation::non-degenerate::cdf-approx} shows the result for deterministic and random designs.
Our method shows clear advantage in accuracy of controlling the empirical coverage probability around the nominal $90\%$ level, significantly improves over normal approximation, especially for small $n$'s.  As $n$ grows large, our method's speed advantage over bootstrap methods becomes clearer.  Compared to inference based on complete U-statistic, our method effectively reduces computational complexity (much flatter computation time curve) without noticeable loss in risk control accuracy.
All methods (except subsampling bootstrap) produce similar CI lengths.  This echoes our earlier remarks that the CI length reflects a different aspect of U-statistic reduction (inference power, Section \ref{subsec::intro::two-aspects-of-trade-off}); and different approaches may perform similarly in this aspect, if they are all asymptotically normal approximations.

\subsection{Simulation 2: degenerate U-statistics}
\label{simulation::dengenerate}

\begin{figure}[h!]
    \centering
    \makebox[\textwidth][c]{
        \raisebox{0.3em}{
    \includegraphics[width=0.26\textwidth]{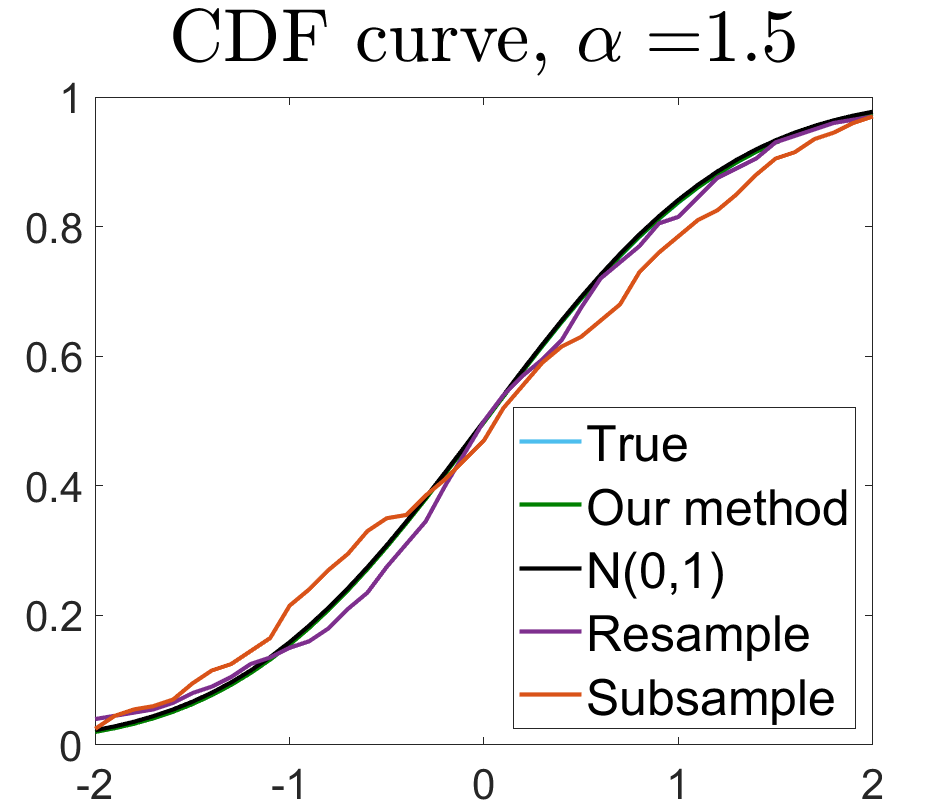}}
    \includegraphics[width=0.25\textwidth]{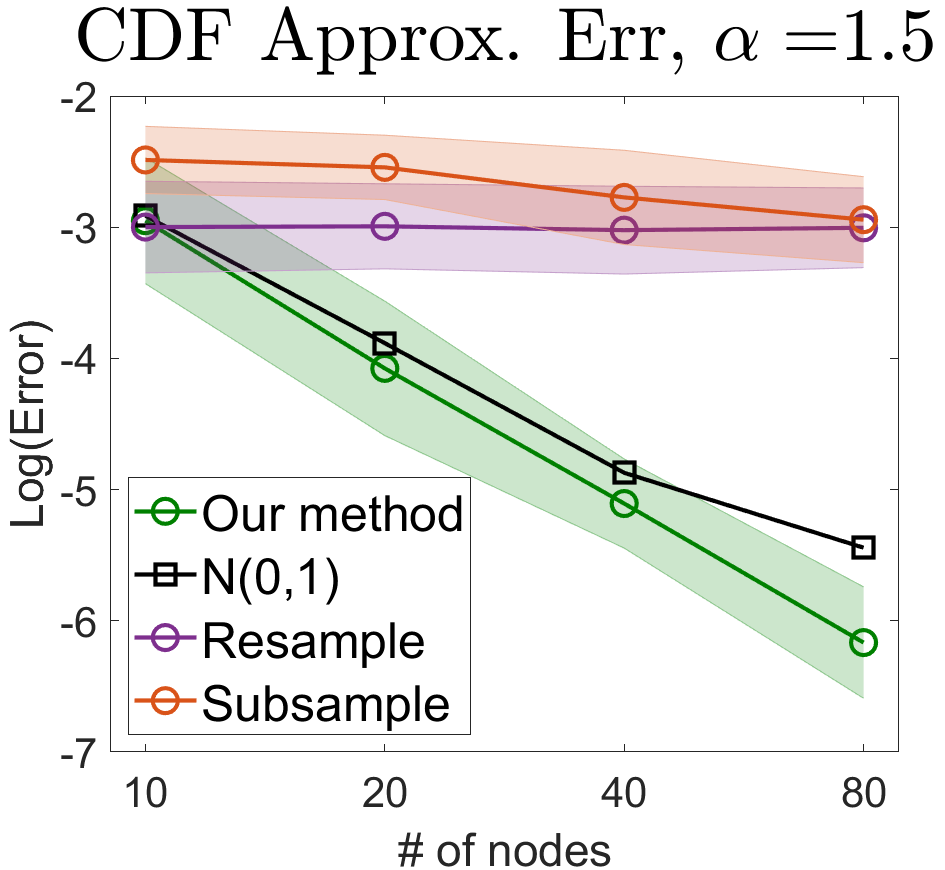}
    \includegraphics[width=0.25\textwidth]{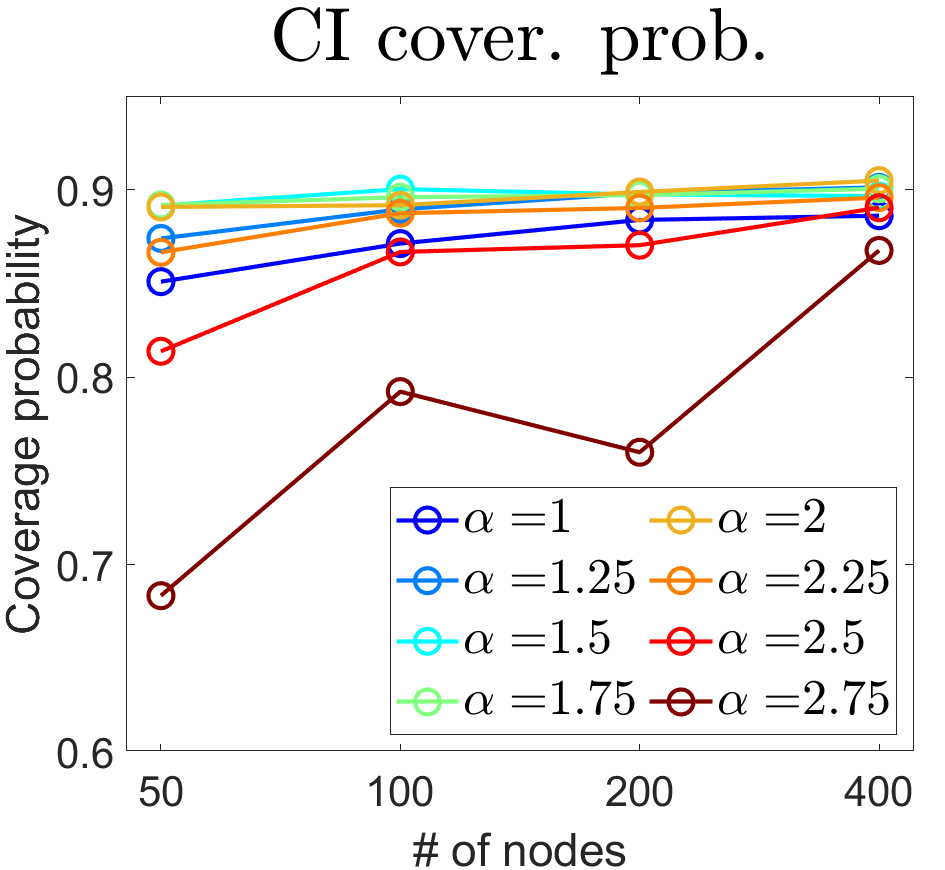} 
    \includegraphics[width=0.25\textwidth]{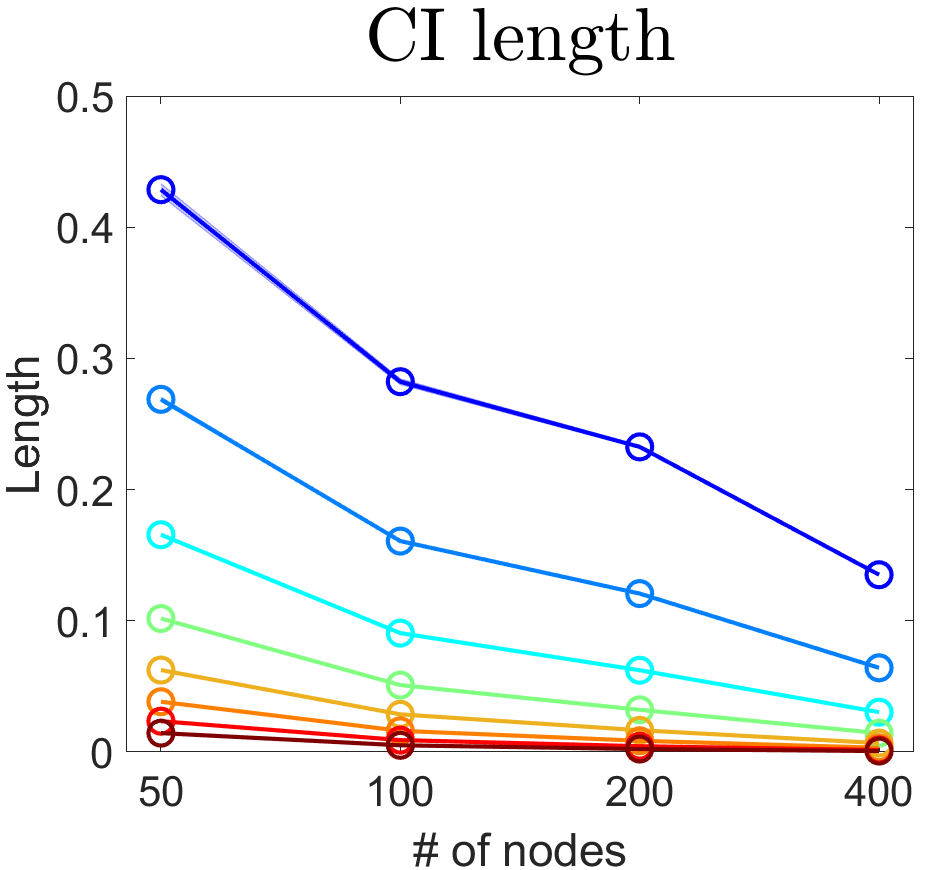}  
    \includegraphics[width=0.25\textwidth]{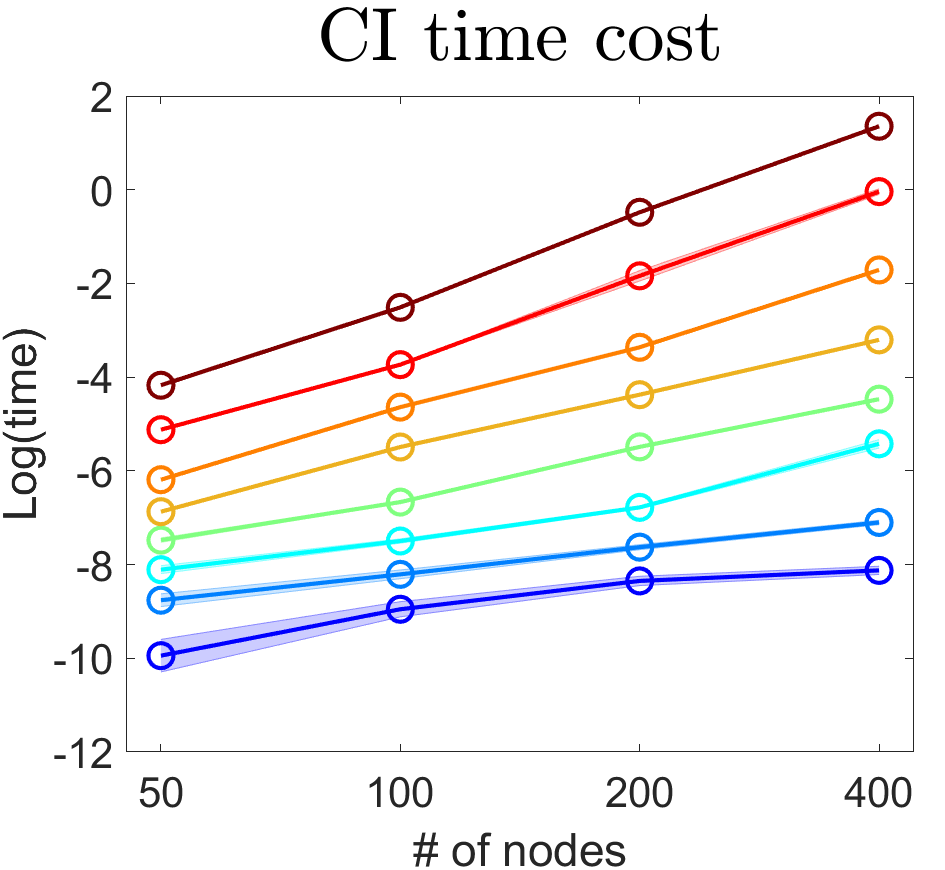}  
    }
    \caption{  
    Degenerate U-statistics:
    Column 1: CDF curves;
    column 2: CDF approximation error;
    columns 3--5: CI performance of our method with various $\alpha$'s.
    }
    \label{simulaion::degenerate cdf curve}
\end{figure}

We generate $X_1,\ldots,X_n\stackrel{\rm i.i.d.}\sim$~Uniform$(0,1)$ and with the kernel $h(x_1, x_2, x_3) := 27\prod_{\ell=1}^3 \big\{\sin(2x_\ell)-\sin^2 1\big\}$. 
We set $n \in \{10, 20, \ldots, 160\}$. $\alpha = 1.5$ and $n_{\rm MC} = 10^6$, and repeat each individual experiment 30 times.
In Figure \ref{simulaion::degenerate cdf curve}, plots 1 and 2 confirm our method's superiority in accurately approximating $F_{\frac{U_J-U_n}{|\jna|^{-1/2}\cdot \tilde \sigma_h}+\delta_J \big|X_{[1:n]}}(u)$.
Next, we study the very interesting observation of our Theorem \ref{main-theorem::one-sample::degenerate::random::normality} and in earlier literature \citet{weber1981incomplete, chen2019randomized} that the randomness and incompleteness of a randomly designed $\jna$ reinstates normality in $U_J$.
To this end, we set $\alpha\in \{1, 1.25, 1.5, \ldots, 2.75\}$, $n\in \{50, 100, 200, 400\}$, and set $n_{\rm MC}=500$ due to the high computational cost at $\alpha=2.75$.
Plot 3 in Figure \ref{simulaion::degenerate cdf curve} confirms the prediction of our Theorem \ref{main-theorem::one-sample::degenerate::random::normality} that the best $\alpha$ in this example should be close to $k_0/3=1$, despite setting $\alpha$ slightly above 1 noticeably improves the accuracy of CI level control, possibly due to numerical reasons; 
whereas setting $\alpha$ close very close to $k_0=3$ (e.g. $\alpha=2.75$) would significantly depreciate the accuracy of our method, since the distribution of $T_J$ at $\alpha=k_0\geq 2$ is non-Gaussian.  The last two plots in Figure \ref{simulaion::degenerate cdf curve} confirm our method's fine performance in CI length (which reflects power) and computation speed.

\subsection{Simulation 3: network moments}
\label{simulation:network-U-stat}

\begin{figure}[h!]
    \centering
    \makebox[\textwidth][c]{
    \includegraphics[width=0.25\textwidth]{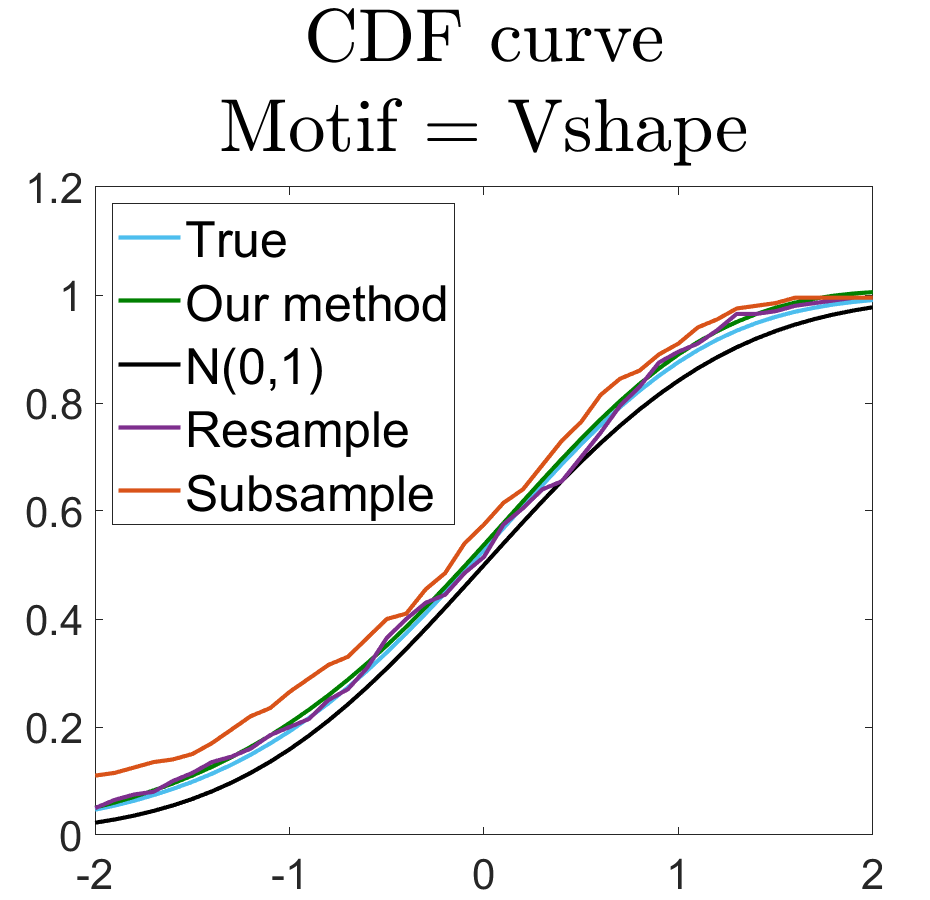} 
    \includegraphics[width=0.25\textwidth]{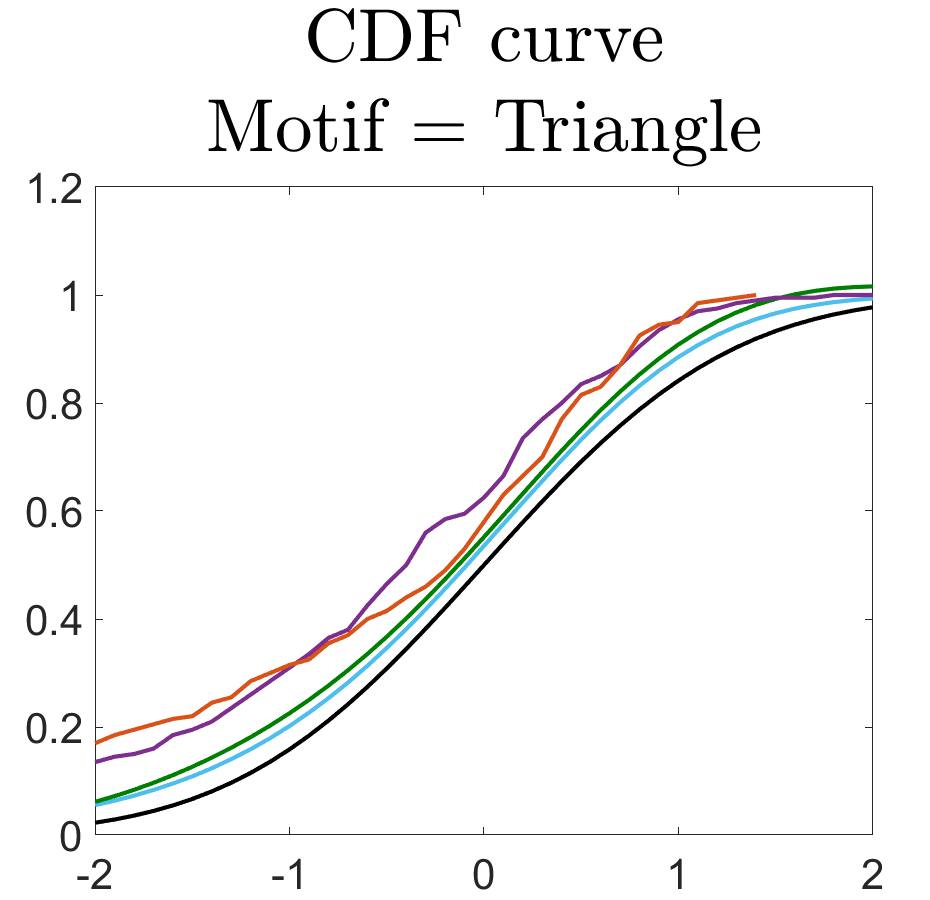} 
    \includegraphics[width=0.25\textwidth]{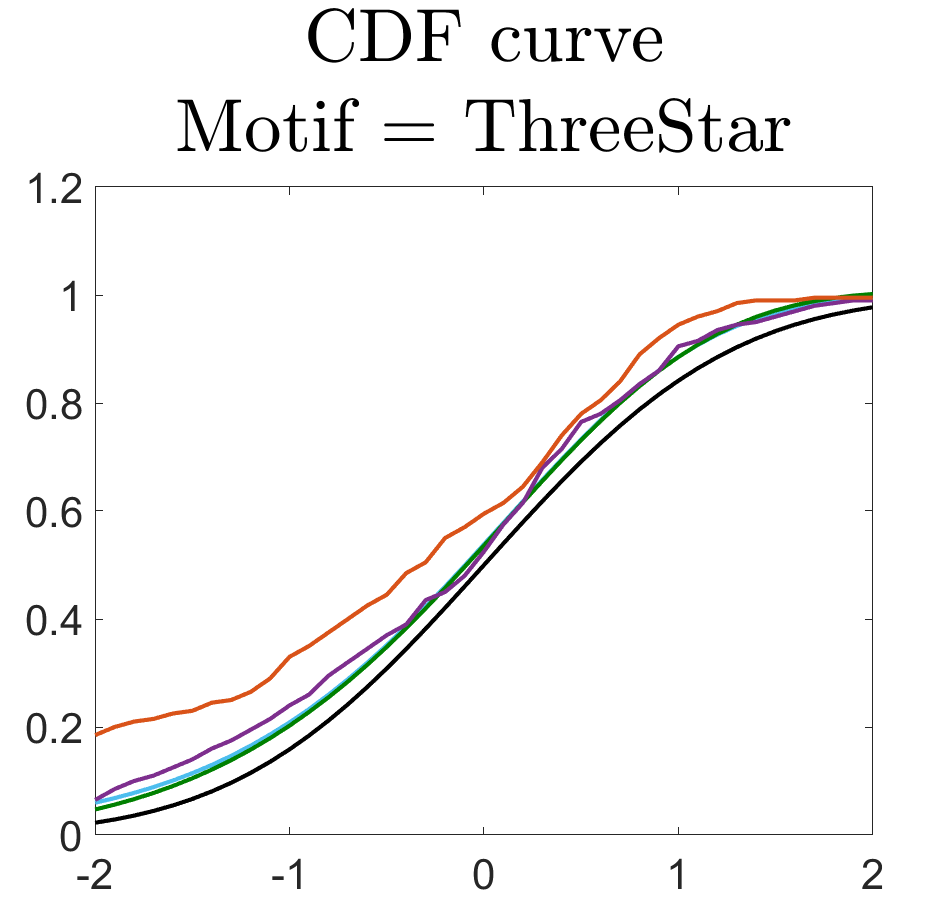} 
    \includegraphics[width=0.25\textwidth]{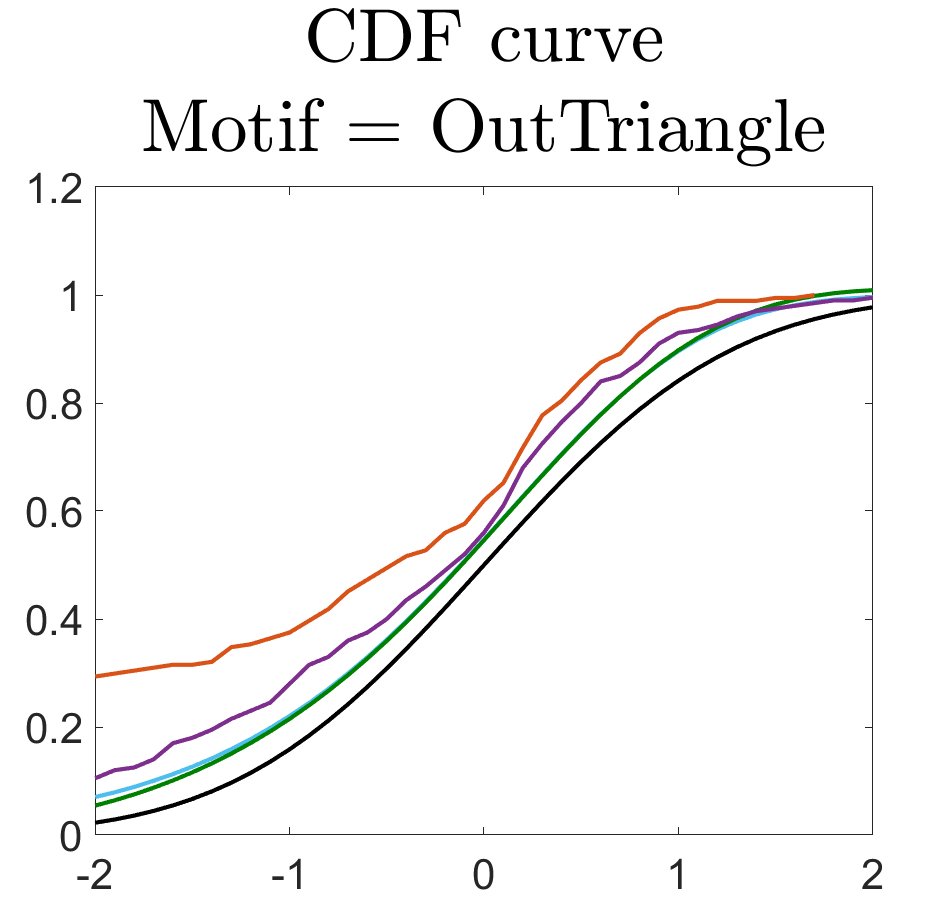} 
    }
    \makebox[\textwidth][c]{
    \includegraphics[width=0.25\textwidth]{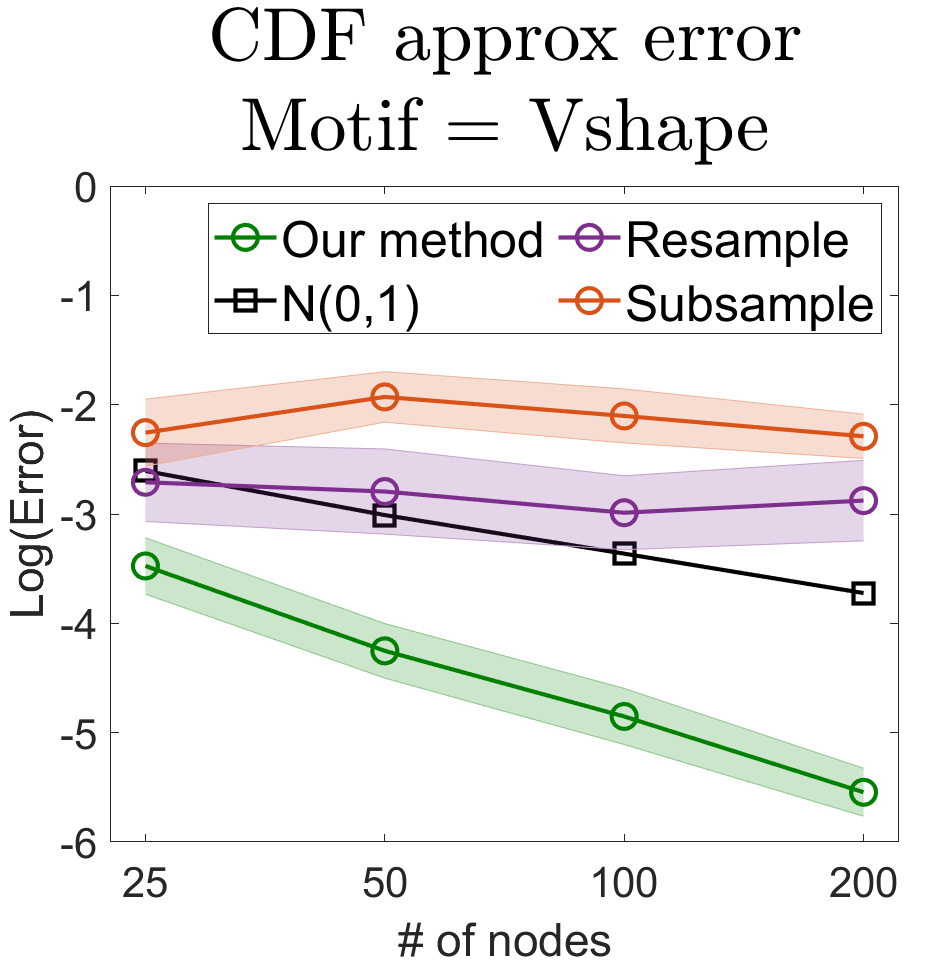} 
    \includegraphics[width=0.25\textwidth]{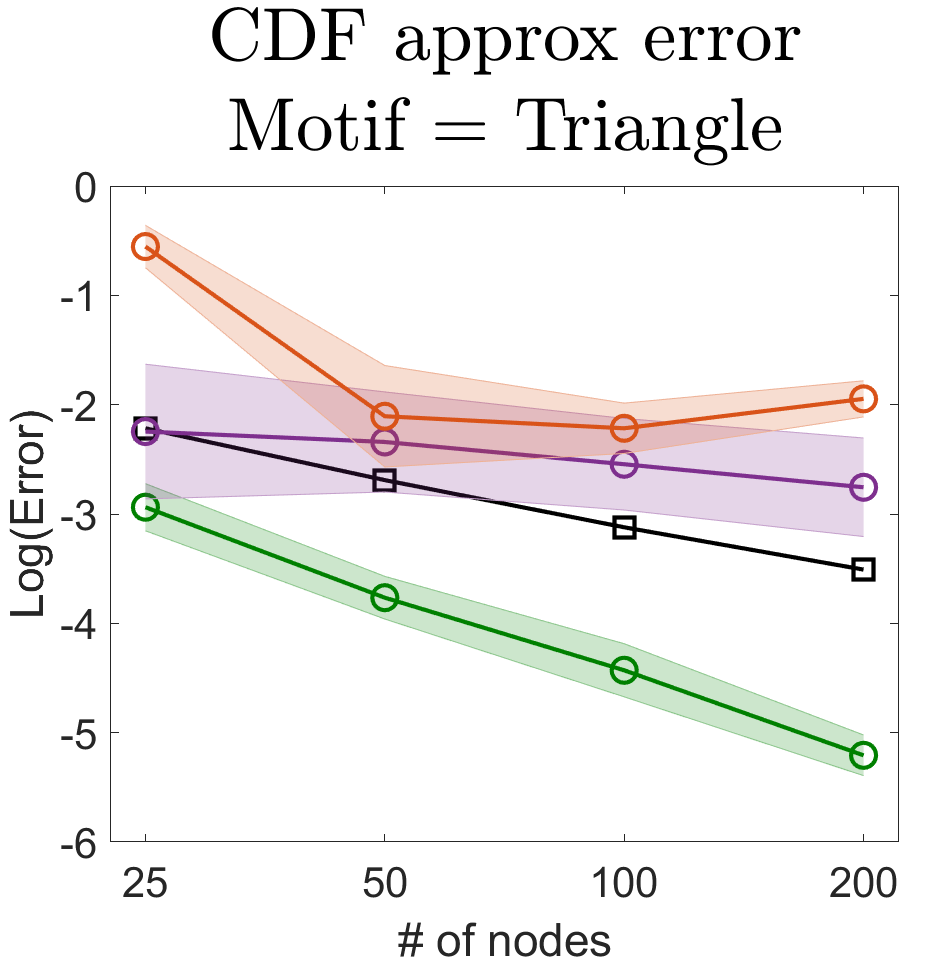}
    \includegraphics[width=0.25\textwidth]{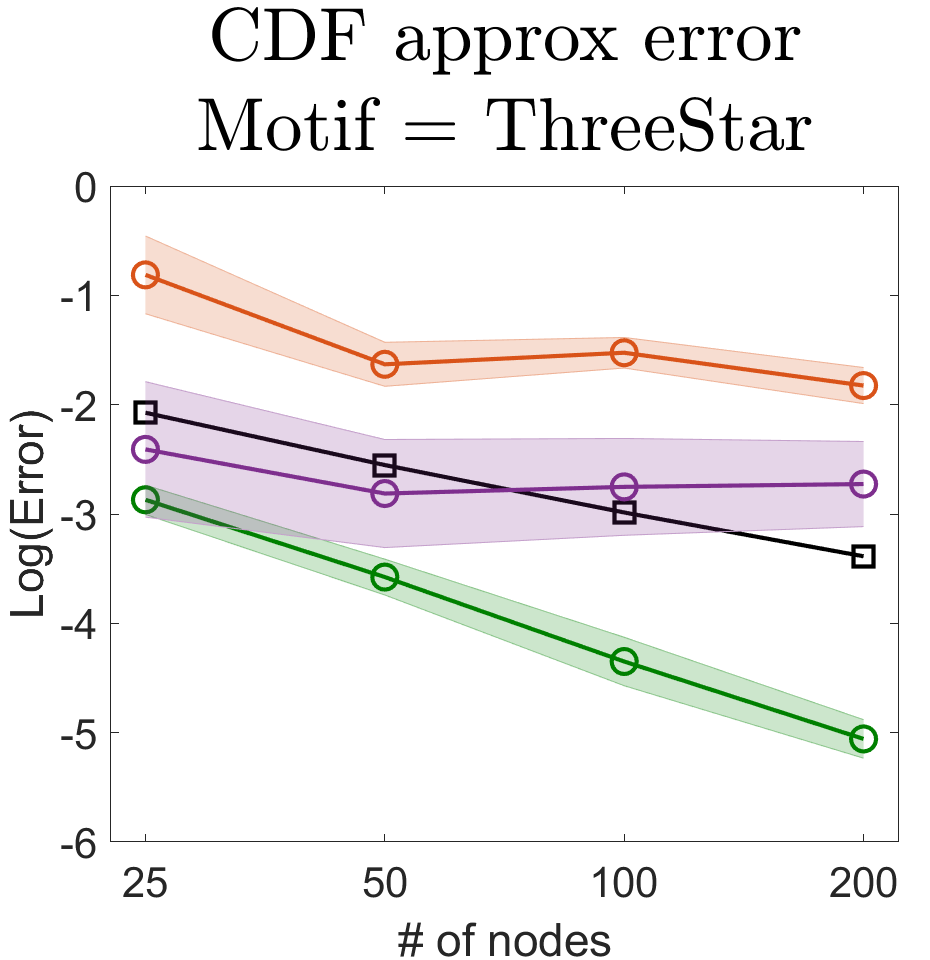} 
    \includegraphics[width=0.25\textwidth]{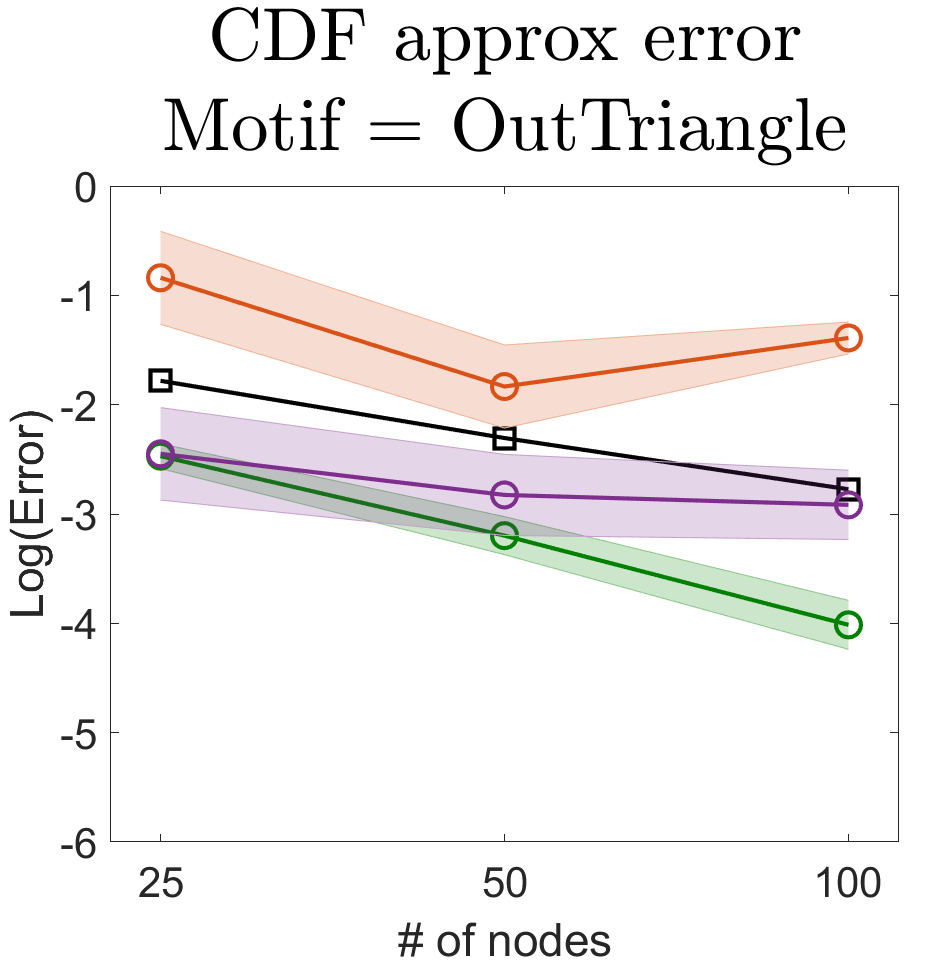} 
    }
    \caption{
    Network moments:
    Row 1: CDF curves, true CDF~$=F_{\hat{T}_J+\delta_J}(u)$,
    $n = 100$.
    Our method: $\alpha=2.5$, random $\jna$ (scheme \ref{jna-choice-1:SRS-w-replace}).
    Row 2: CDF approximation errors\protect\footnotemark.
    }
    \label{simulations::network::cdf-approx-error}
\end{figure}

We first compare the CDF approximation accuracy of different methods for dense networks generated by the ``{\tt BlockModel}'' in \citet{zhang2020edgeworth} with $\rho_n \asymp 1$.
We experimented the four graphons (Edge, V-shape, Triangle and Three-Star) in \citet{zhang2020edgeworth}, plus a new and more complex graphon, which we call ``Out-Triangle'' \citep{biemann2012quantifying}, defined by its edge set $\{(1,2),(2,3),(1,3),(1,4)\}$.
By Theorem \ref{theorem::main-theorem::network-U-stat}, $R=$~Edge follows a different guideline for choosing $\alpha$.  Due to page limit, we sink the study of the edge motif to Supplementary Material.
For the other four motifs, we set $\alpha=2.5$, $n \in \{25, 50, 100, 200\}$, and inherit other simulation settings from Section \ref{simulation::simulation-1}.
Figure \ref{simulations::network::cdf-approx-error} shows that our method maintains stable advantage over competing methods in CDF approximation accuracy.
Next, Table \ref{simulation::network::coverage probability in other motifs} presents detailed numerical comparison of the CI's produced by different methods.  We see that our method achieves a clear acceleration over the complete U-statistic method with only slightly depreciated confidence level control.  
Our method's speed is competitive even compared to the normal approximation.  
It is significantly faster and more accurate than subsampling bootstrap, and much faster than resampling bootstrap of comparable accuracy.

\footnotetext{The Monte Carlo evaluation of the true CDF timed out for Out-Triangle when $n=200$.}

\begin{table}[ht!]
\centering
 \caption{Network moments: performance of $95\%$ CI's, $n = 100$, BlockModel, Mean(SD)\\}
\label{simulation::network::coverage probability in other motifs}
\begin{adjustbox}{center, max width=0.85\textwidth}
\begin{tabular}{c|ccccc}\hline
Motif  & V-shape & Triangle & ThreeStar & OutTriangle\\\hline
\bigcell{c}{Our method \\($\alpha=2.5$)\\($n_{\rm MC}=10^4)$}
& \bigcell{c}{Coverage~$=0.952(0.215)$\\Length~$=0.173(0.025)$\\LogTime~$=-4.390(0.013)$} & \bigcell{c}{$0.949(0.219)$\\$0.035(0.006)$\\$-4.604(0.035)$} & \bigcell{c}{$0.945(0.228)$\\$0.124(0.024)$\\$-3.893(0.015)$} & \bigcell{c}{$0.940(0.238)$\\$0.200(0.049)$\\$-2.919(0.020)$}\\\hline
\bigcell{c}{Norm. Approx. \\($\alpha=2.5$)\\($n_{\rm MC}=10^4)$}
&
\bigcell{c}{$0.939(0.238)$\\$0.173(0.025)$\\$-4.549(0.072)$} & \bigcell{c}{$0.934(0.249)$\\$0.035(0.006)$\\$-4.869(0.019)$} & \bigcell{c}{$0.930(0.255)$\\$0.124(0.024)$\\$-4.062(0.018)$} & \bigcell{c}{$0.922(0.268)$\\$0.200(0.049)$\\$-3.058(0.010)$}\\\hline
\bigcell{c}{Complete\\($n_{\rm MC}=500)$} & \bigcell{c}{$0.953(0.211)$\\$0.172(0.025)$\\$-4.114(0.078)$} & \bigcell{c}{$0.951(0.215)$\\$0.034(0.006)$\\$-4.522(0.018)$} & \bigcell{c}{$0.948(0.223)$\\$0.123(0.024)$\\$-0.727(0.016)$} & \bigcell{c}{$0.942(0.234)$\\$0.199(0.048)$\\$0.272(0.026)$}\\\hline
\bigcell{c}{Resample\\($\alpha=2.5$)\\($n_{\rm MC}=500)$} & \bigcell{c}{$0.947(0.224)$\\$0.179(0.030)$\\$0.756(0.010)$} & \bigcell{c}{$0.952(0.214)$\\$0.038(0.007)$\\$0.559(0.014)$} & \bigcell{c}{$0.947(0.223)$\\$0.131(0.028)$\\$1.266(0.012)$} & \bigcell{c}{$0.955(0.208)$\\$0.228(0.057)$\\$2.356(0.012)$}\\\hline
\bigcell{c}{Subsample\\($\alpha=2.5$)\\($n_{\rm MC}=500)$} & \bigcell{c}{$0.964(0.187)$\\$0.348(0.082)$\\$-2.716(0.015)$} & \bigcell{c}{$0.923(0.267)$\\$0.071(0.026)$\\$-2.791(0.014)$} & \bigcell{c}{$0.959(0.198)$\\$0.548(0.193)$\\$-2.301(0.025)$} & \bigcell{c}{$0.929(0.256)$\\$0.979(0.445)$\\$-1.787(0.014)$}\\\hline
\end{tabular}
\end{adjustbox}
\end{table}

Our last experiment numerically pokes the asymptotic normality of $\hat T_J'$ for sparse networks under different amounts of computation reduction.
Here, we set $R=$~Triangle and the graphon to be {\tt BlockModel}.
We tested two sparsity regimes: $\rho_n\asymp n^{-1/2}$ and $n^{-2/3}$.
Our Theorem \ref{theorem::main-theorem::network-normality} predicts $\hat T_J'\to N(0,1)$ when $\rho_n^s n^{\alpha-1}\to 0$ and $\rho_n^s n^\alpha\to \infty$, reflected by our simulation settings $\alpha=2$ for $\rho_n\asymp n^{-1/2}$ and $\alpha=2.5$ for $\rho_n\asymp n^{-2/3}$, respectively.  For each $\rho_n$, we also tested smaller $\alpha$'s.
We repeated each configuration $n_{\rm MC} = 500$ times and compared the sampling distribution of $\hat T_J'$ to $N(0,1)$.
The QQ plots in Figure \ref{theorem::main-theorem::network-normality} well-matches our Theorem \ref{theorem::main-theorem::network-normality}'s prediction.  In plots 2 \& 4, we clearly see the expected convergence to $N(0,1)$.  In contrast, in plots 1 \& 3, where $\rho_n^s n^\alpha\to 0$, we see non-Gaussian behaviors, suggesting that the conditions of our Theorem \ref{theorem::main-theorem::network-normality} seem minimal.

\section{Data examples}
\label{section::data-examples}

\subsection{Data example 1: Stock market data}
\label{data-example::stock-market}
\begin{figure}[h!]
    \centering
    \makebox[\textwidth][c]{
    \includegraphics[width=0.5\textwidth]{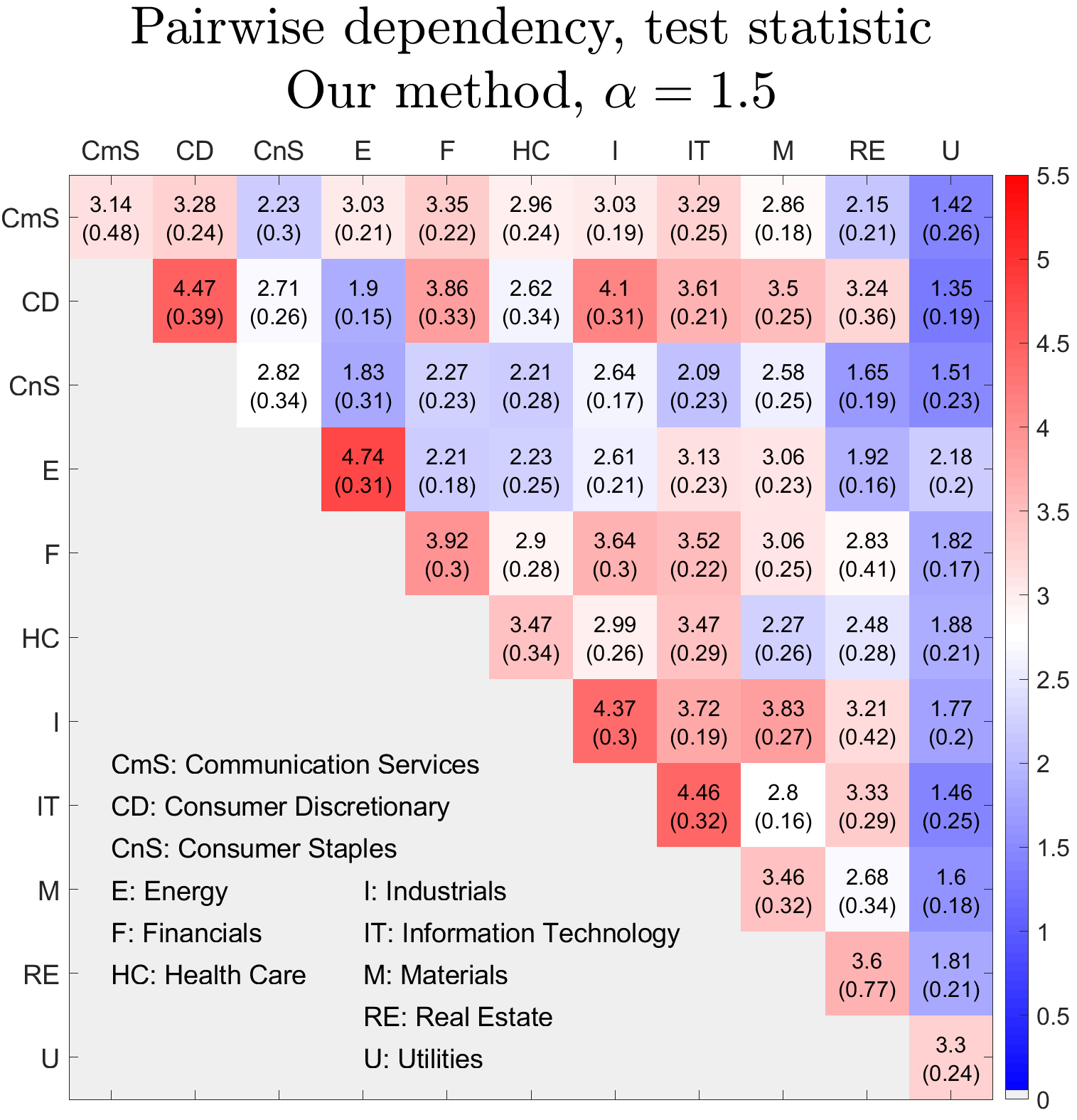} 
    \includegraphics[width=0.5\textwidth]{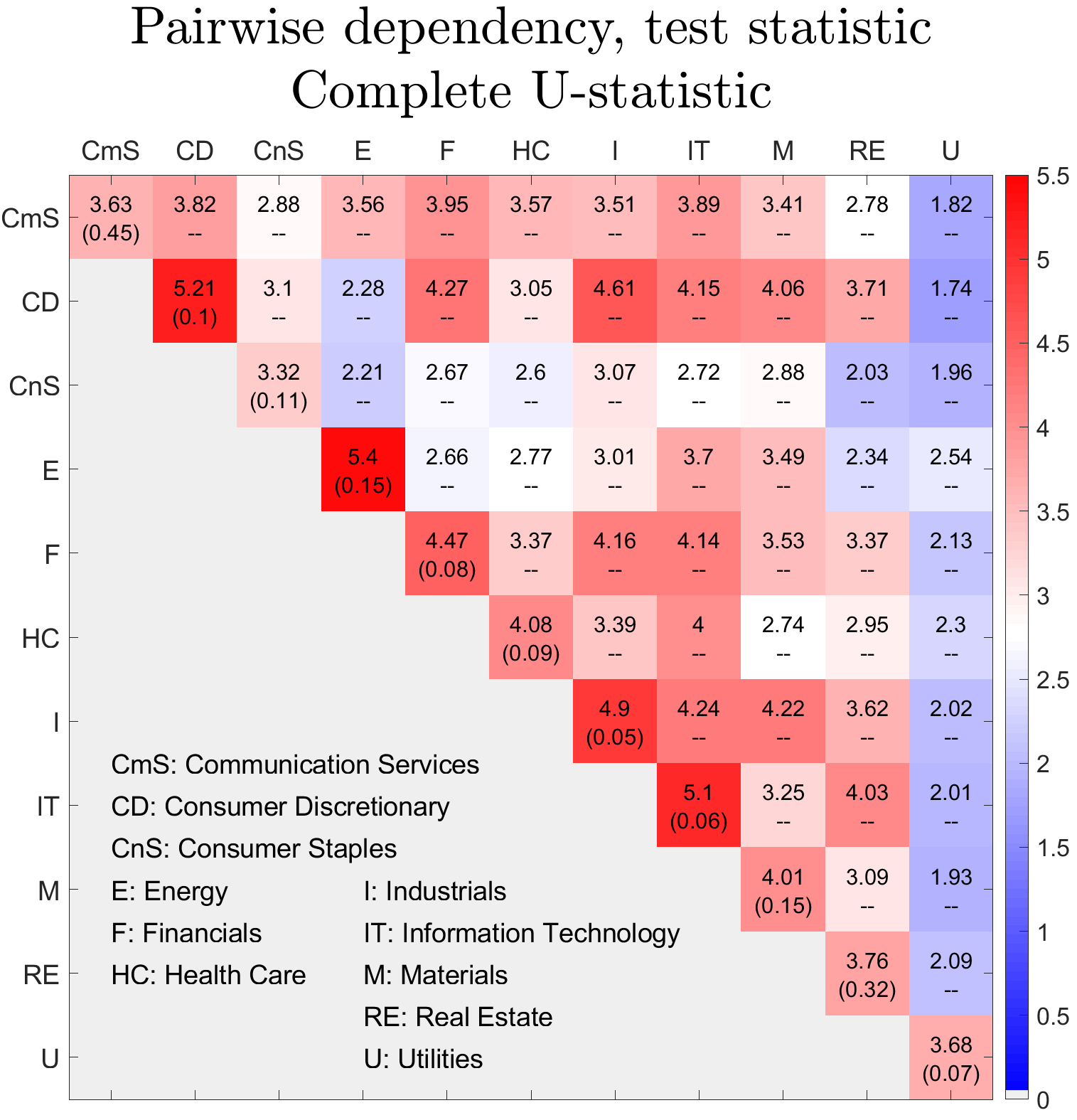} 
    }
    \caption{
    Pairwise dependency test:  heatmaps of test statistics.  High values (red): high detected dependency.
    Each cell reports mean(std.) of test statistics over 30 repeated experiments, except the off-diagonal of complete U-statistic method (no repetition needed).
    }
    \label{fig::data-example::stockmarket::heatmap}
    
\end{figure}

The S\&P 500 historical data \citep{sp500} records the daily prices of 412 stocks from 11 sectors.
Following \citet{chakraborty2021new}, we computed the \emph{monthly logarithmic return rates} of each stock from 1-March-2000 to 29-August-2022, yielding $n=138$ observations.
The goal is to assess the pairwise dependency between all sectors using an independence test.
Denote the stock returns from a sector $X$ containing $p$ stocks as $S_t^{X} = (S_{it}^X,...,S_{pt}^X)$, $t\in [1:138]$; similarly define $S_t^Y$.
We measure dependency by dCov, rewritten as a U-statistic (Lemma 1 of \citet{yao2018testing}).
\begin{align}
    {\rm dCov}^2(X,Y) 
    :=&~
    \binom {n}{4}^{-1}
    \sum_{i<j<q<r} 
    h(Z_i, Z_j, Z_q, Z_r),
\end{align}
where
$
    h(Z_i, Z_j, Z_q, Z_r) := 
    \sum_{s,t,u,v}^{i,j,q,r}
    (a_{st}b_{uv}+a_{st}b_{st} - a_{st}b_{su}-a_{st}b_{tv})
    \big/ 24
$
\footnote{The summation notation ``$\sum_{s,t,u,v}^{i,j,q,r}$'' means $\sum_{(s,t,u,v)\in{\rm All.Permutations}(i,j,q,r)}$.}, 
$a_{ij} = \|S_i^{X} - S_j^{X}\|_2$, and $b_{ij} = \|S_i^{Y} - S_j^{Y}\|_2$. 
We set $\alpha=1.5$ and used it to evaluate the test statistic for $H_0: \ep[{\rm dCov}^2]=0$ between each sector pair.  
Observing that $U_J$'s are on the same order as $\hat \xi_1$, we used our formula for the non-degenerate case.
As a reference, on the diagonal, we also randomly split the stocks in each sector into two sets and evaluated their dependency using the same method. 
Figure \ref{fig::data-example::stockmarket::heatmap} shows that our inference procedure for reduced U-statistics well-preserves the result of the complete U-statistic but computes much faster (Table \ref{tab::data-examples::time-cost-comparison}).
On the diagonal, the sectors that exhibit strongest inner dependency include \emph{CD, E, F, I} and \emph{IT}.  This is understandable since they tend to be more sensitive to global economic changes.  In contrast, members of \emph{CmS, CnS} and \emph{U} sectors focus more on local markets, so their price fluctuations are less synchronized.
This understanding also applies to cross-sector relations, such as the tight connection between the pairs (\emph{CD}, \emph{I}) and (\emph{I}, \emph{IT}), whereas \emph{U} is comparatively less dependent on other sectors except \emph{E}.

\subsection{Data example 2: UCR time series data (Earthquakes, Starlight)}
\label{data-example::UCR-time-series}
\begin{figure}[h!]
    \centering
    \makebox[\textwidth][c]{
    \raisebox{0.2em}{
    \includegraphics[width=0.26\textwidth]{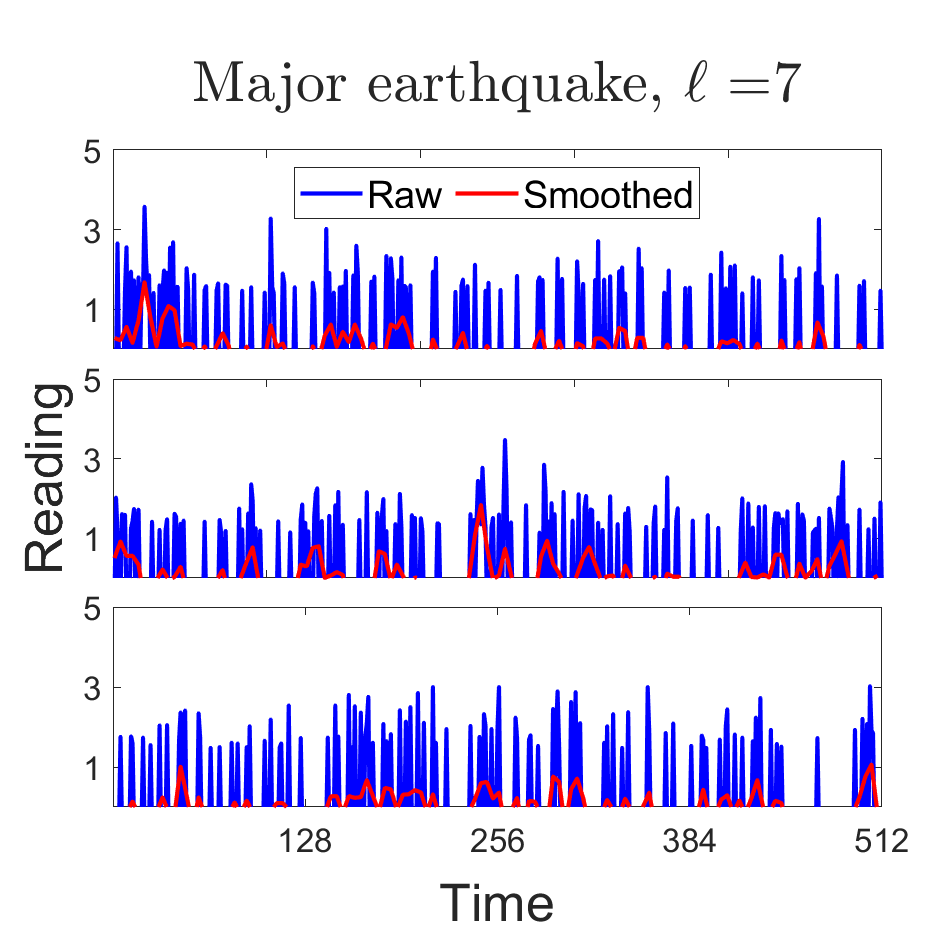} 
    \includegraphics[width=0.26\textwidth]{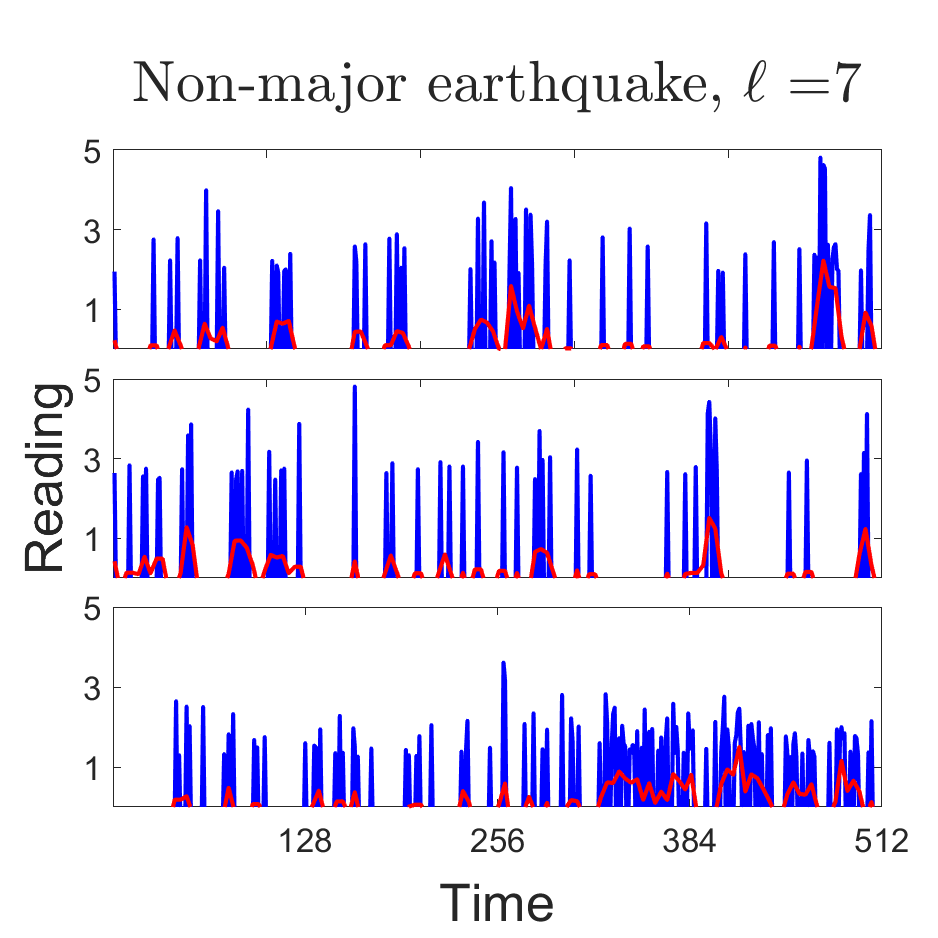} }
    \includegraphics[width=0.26\textwidth]{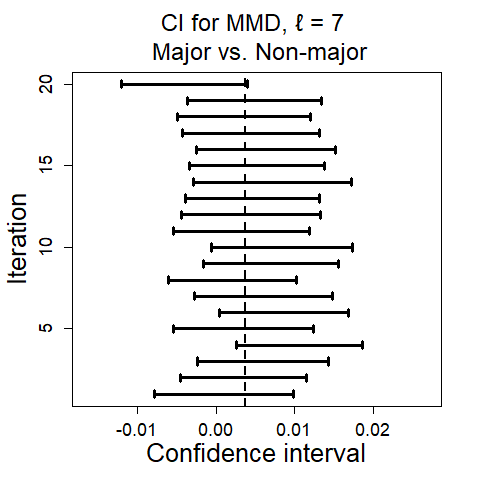}
    \includegraphics[width=0.26\textwidth]{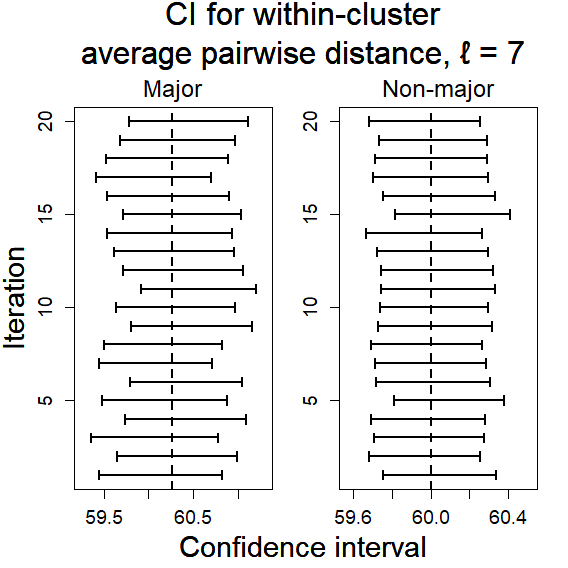} 
    }\\
    \makebox[\textwidth][c]{
    \raisebox{0.2em}{
    \includegraphics[width=0.26\textwidth]{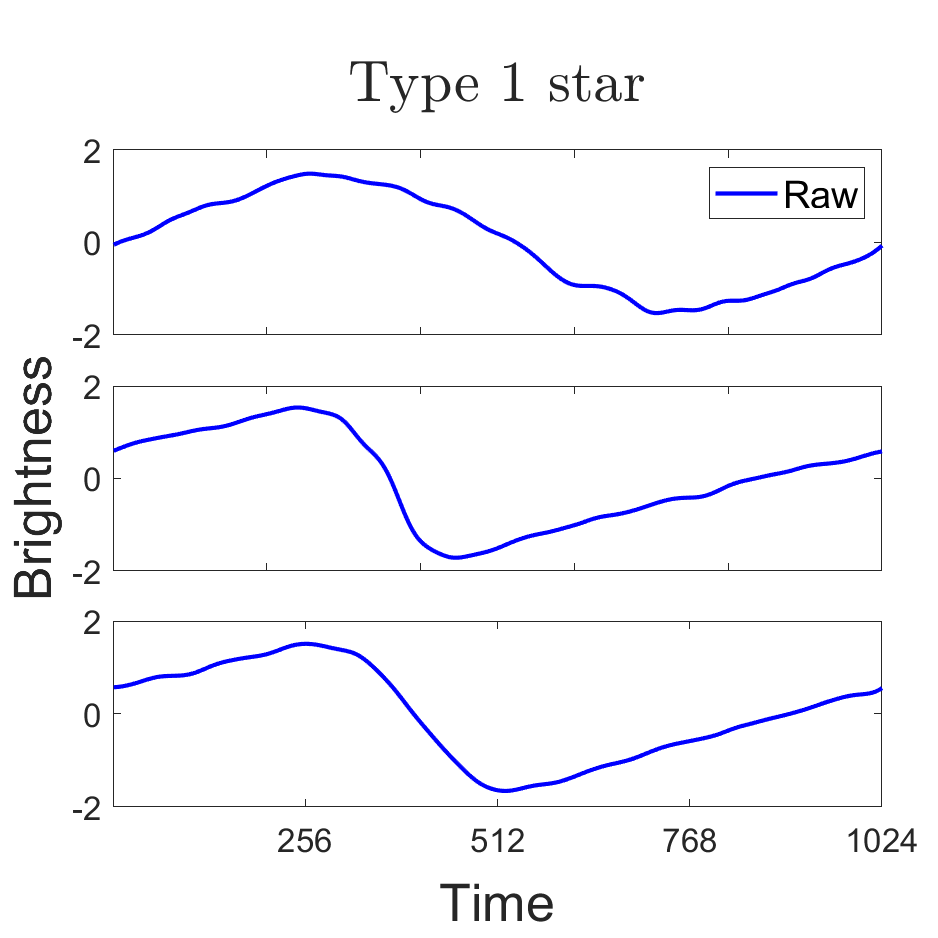} 
    \includegraphics[width=0.26\textwidth]{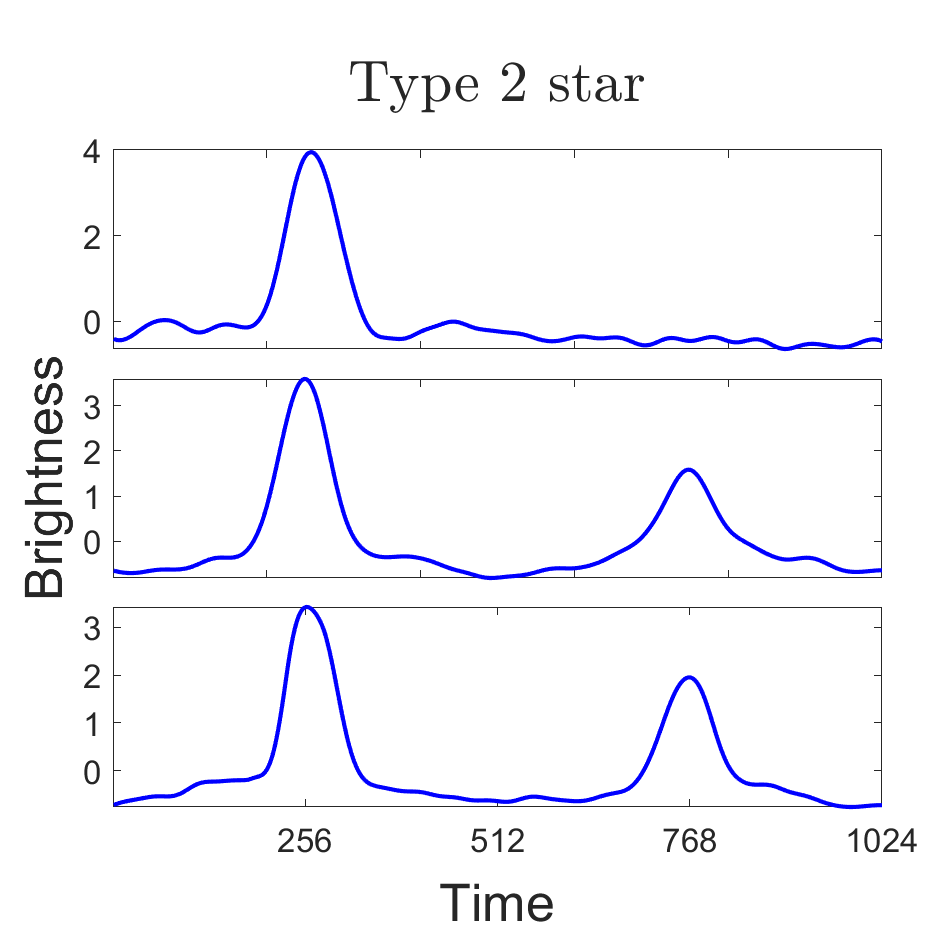} }
    \includegraphics[width=0.26\textwidth]{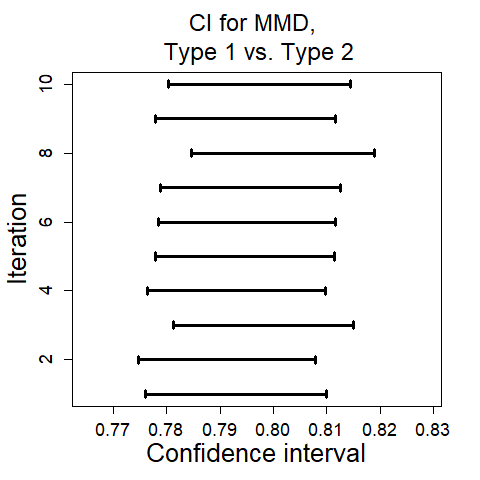} 
    \includegraphics[width=0.26\textwidth]{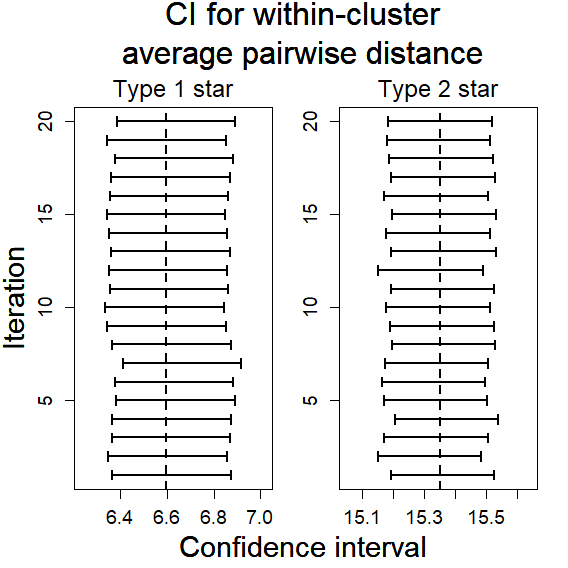} 
    }
    \caption{
        Results of data example 2. 
        Row 1: \emph{Earthquakes}; row 2: \emph{Starlight}. 
        Column 1 \& 2: raw and pre-pocessed curves; 
        column 3: 90\% CI of based on reduced between-cluster MMD; 
        column 4:  90\% CI of within-cluster average pairwise distance (using SRVF \citep{srivastava2011registration}).
        Dashed line: complete U-statistic (evaluations of complete U-statistics timed out ($>48$ hours) in most settings for \emph{Starlight} data).
    }
    \label{fig::data-example::UCR::confidence interval}
\end{figure}

In the second example, we analyze two UCR time series data sets \citep{UCRArchive2018}: \emph{Earthquakes} and \emph{Starlight}.
The former records $n=461$ earthquake curves of length $T=512$.
All curves are classified into $K=2$ types: $n_0=368$ non-major and $n_1=93$ major ones.
Following \citet{chakraborty2021new} and \citet{zhu2021interpoint}, we treat each earthquake curve as a data point in a Hilbert space and aim at comparing the population distributions of the curves of different types using \emph{Maximum Mean Discrepancy (MMD)} metric.
We measure the distance between two earthquake curves by comparing their SRVF transforms \citep{srivastava2011registration} -- a method that synchronizes the two curves' phases in presence of amplitude discrepancy.
The computation of SRVF for each curve pair is slow \citep{strait2019automatic},
so we reduced each curve $\{x_t\}_{t=1}^{512}$ into $\big\{\tilde x_t:={\rm Mean}\big( x_{[\{t-(\ell-1)/2\}:\{t+(\ell-1)/2\}]}\big)\big\}_{t\in\{4k+1, k\in[0:127]\}}$.
Due to page limit, here we only present $\ell=7$; results for more window sizes can be found in Supplementary Material.
We set $\alpha=1.5$ and apply our method to reduce the MMD, rewritten as a U-statistic according to Equation (6) in \citet{schrab2021mmd} with the RBF kernel $k(x,y):=\exp(-{\rm SRVF}(x,y)^2/5000)$.
To apply this MMD formula, we subsampled the larger class (small earthquakes) to equate the two sample sizes.
We also computed the averaged pairwise distance (using SRVF) within each cluster to illustrate its inner cohesion.
Row 1 of Figure \ref{fig::data-example::UCR::confidence interval} shows the result.
We certainly do not have access to the population MMD value, but deeming the empirical (complete) two-sample MMD as a good approximation, our Cornish-Fisher confidence intervals with randomized design \ref{jna-choice-1:SRS-w-replace} shows good coverage.
The same fine performance shows up for the within-cluster average pairwise distances.

Next, we apply the same analysis to the much larger \emph{Starlight} data set that contains 3 types of stars ($n_1=1329$, $n_2=2580$ and $n_3=5327$).
Each curve is recorded as a length $1024$ sequence.
We down-sampled each curve sequence to length $128$ without smoothing, because the starlight curves are much smoother than the earthquake curves.
Even so, evaluating a complete U-statistic for comparing any two star types remains computationally infeasible, due to the large sample sizes.
Our method with $\alpha=1.5$ makes it possible to implement a reduced version of Equation (6) in \citet{schrab2021mmd} with the RBF kernel $k(x,y):=\exp(-{\rm SRVF}(x,y)^2/100)$.
Due to page limit, in row 2 of Figure \ref{fig::data-example::UCR::confidence interval}, we only present the result for the comparison between type 1 and type 2 stars.
Compared to the \emph{Earthquakes} data, the MMD CI's for \emph{Starlight} are much stabler, understandably due to the latter's much larger sample sizes of each cluster.
Also, some MMD CI's of the \emph{Earthquakes} data contain 0 (will not reject $H_0$), while all CI's for the \emph{Starlight} data clearly support a two-sided alternative.
This is not surprising, given the much smaller within-cluster distance (column 4) and the clearer between-cluster differences (compare columns 1 \& 2) in the latter data.

\begin{table}[h!]
\centering
\caption{Time cost: our method ($\alpha = 1.5$) vs. complete U-statistic}
\label{tab::data-examples::time-cost-comparison}
\begin{adjustbox}{center, max width=0.8\textwidth}
\begin{tabu}[t]{ccccccccccccccccccccccccccccccccccc}
\tabucline[1.5pt]{-}

\multicolumn{7}{c|[1.5pt]}{Time cost} &
\multicolumn{7}{c|[1.5pt]}{Stock Market ($r=4$) }
& \multicolumn{21}{c}{Earthquakes ($r=2$)} \\
\cline{8-35}
\multicolumn{7}{c|[1.5pt]}{(Unit = sec.)}& \multicolumn{7}{c|[1.5pt]}{All} & 
\multicolumn{7}{c}{Major} & 
\multicolumn{7}{c|}{Non-major} & 
\multicolumn{7}{c}{Maj. vs. Non-Maj.}  \\
\tabucline[1.5pt]{-}
\multicolumn{7}{c|[1.5pt]}{Our method} &
\multicolumn{7}{c|[1.5pt]}{3.47}&
\multicolumn{7}{c}{303.94}&
\multicolumn{7}{c|}{2471.70}&
\multicolumn{7}{c}{1223.50}\\

\multicolumn{7}{c|[1.5pt]}{Complete U} &
\multicolumn{7}{c|[1.5pt]}{8099.73}&
\multicolumn{7}{c}{708.99}&
\multicolumn{7}{c|}{11199.92}&
\multicolumn{7}{c}{17912.91}\\
\tabucline[1.5pt]{-}

\multicolumn{5}{c|[1.5pt]}{Time cost} &
\multicolumn{30}{c}{Starlight ($r=2$)} \\
\cline{6-35}

\multicolumn{5}{c|[1.5pt]}{(Unit = sec.)}&
\multicolumn{5}{c}{Type 1} & 
\multicolumn{5}{c}{Type 2} & 
\multicolumn{5}{c|}{Type 3} & 
\multicolumn{5}{c}{1 vs. 2} & 
\multicolumn{5}{c}{1 vs. 3} &
\multicolumn{5}{c}{2 vs. 3}  \\
\tabucline[1.5pt]{-}

\multicolumn{5}{c|[1.5pt]}{Our method} &
\multicolumn{5}{c}{4512.95}&
\multicolumn{5}{c}{12773.76}& 
\multicolumn{5}{c|}{41282.26} &
\multicolumn{5}{c}{19140.13}& \multicolumn{5}{c}{19149.33}&
\multicolumn{5}{c}{50413.75}\\

\multicolumn{5}{c|[1.5pt]}{Complete U} &
\multicolumn{5}{c}{48227.72}&
\multicolumn{5}{c}{158233.7}&
\multicolumn{5}{c|}{(Time out)}&
\multicolumn{5}{c}{(Time out)}&
\multicolumn{5}{c}{(Time out)}&
\multicolumn{5}{c}{(Time out)}\\
\tabucline[1.5pt]{-}

\end{tabu}
\end{adjustbox}
\end{table}%

\section{Discussion}
\label{section::discussion}

Our work assumes that the degree of degeneracy $k_0$ is known.
This is a common practice in the analysis of degenerate U-statistics and seemingly cannot be circumvented by bootstraps: as Remark 3.2 in \citet{chen2019randomized} pointed out, the knowledge of $k_0$ is critical for correctly setting how many $r$-tuples to be randomly sampled.
Fortunately, our proofs imply that each $\xi_k$'s can be estimated by $\hat\xi_k$ defined in Section \ref{section::our-method} with concentration bound $|\hat\xi_k - \xi_k| = \tOp(n^{-1/2}\log^{k/2}n)$.
Therefore, one can simply test $H_0(k):\xi_k=0$ versus $H_a(k):\xi_k={\rm constant}>0$ by comparing $\hat\xi_k$ to $n^{-c_0}$, $c_0\in(0,1/2)$ to find $k_0$.
Another approach, called the ``blocking trick'', by \citet{ho2006two}, breaks the data $X_{[1:n]}$ into $K_0$ blocks $X_{[\{(k-1)(n/K_0)+1\}:(kn/K_0)]}, k\in[1:K_0]$, computes a complete U-statistic on each block, and makes inference using a CLT for the sample mean of $K_0$ i.i.d. terms.
Its unification of degenerate and non-degenerate cases and automatic adaptation to $k_0$, however, comes at the price of heavily wasting data.

Our study throughout this paper exclusively focuses on \emph{data-oblivious} reduction schemes.  Recently, \citet{kong2020design} proposed a \emph{data-aware} reduction scheme, based on their key observation that $X_{[1:r]}\approx Y_{[1:r]}$ implies $h(X_{[1:r]})\approx h(Y_{[1:r]})$, thus by clustering $X_i$'s, one can effectively reduce the U-statistic's computation.
While their method shows very attractive performance, finite-sample higher-order analysis for their method poses an interesting open challenge.
There is also a computational price for \emph{being data-aware}.
For example, in the setting considered by \citet{moon2022interpoint}, the clustering of all $X_i$'s in some Banach space requires computing at least $O(n^2)$ many potentially expensive (like in our second data example) pairwise distances.

\spacingset{1}
\bibliographystyle{chicago}
\bibliography{all-ref}

\end{document}